\documentclass{aa}

\usepackage{graphicx}
\usepackage[varg]{txfonts}
\usepackage{amsmath}
\usepackage{booktabs}
\usepackage{hyperref}
\usepackage{upgreek}

\begin{document}

   \title{Closing the gap: Follow-up observations of peculiar dusty objects close to Sgr~A* using ERIS}

   \titlerunning{ERIS observations of the S cluster}

	\author{F. Pei{\ss}ker\inst{\ref{inst1},\ref{inst2}}
        \and M. Zaja\v{c}ek\inst{\ref{inst3},\ref{inst1}}
        \and V. Karas\inst{\ref{inst6}}
        \and V. Pavl\'{\i}k\inst{\ref{inst6},\ref{inst7}}
        \and E. Bordier\inst{\ref{inst1}}
        \and L. \v{S}ubr\inst{\ref{inst4}}
        \and J. Haas\inst{\ref{inst4}}
        \and M. Melamed\inst{\ref{inst1},\ref{inst5}}
        \and L.~Gro{\ss}ekath{\"o}fer\inst{\ref{inst1}}
        \and N. Schm{\"o}kel\inst{\ref{inst1}}
        \and M. Singhal\inst{\ref{inst4}}      
     } 

	\institute{I.Physikalisches Institut der Universit\"at zu K\"oln, Z\"ulpicher Str. 77, 50937 K\"oln, Germany\label{inst1} \and
		\email{peissker@ph1.uni-koeln.de}\label{inst2}
        \and Department of Theoretical Physics and Astrophysics, Faculty of Science, Masaryk University, Kotlá\v{r}ská 2, 611 37 Brno, Czech Republic\label{inst3}
        \and Astronomical Institute, Czech Academy of Sciences, Bo\v{c}n\'{i} II 1401, CZ-14100 Prague, Czech Republic\label{inst6}
        \and Indiana University, Department of Astronomy, Swain Hall West, 727 E 3$^\text{rd}$ Street, Bloomington, IN 47405, USA \label{inst7}
        \and Charles University, Faculty of Mathematics and Physics, Astronomical Institute, V Hole\v{s}ovi\v{c}k\'ach 2, CZ-18000 Prague, Czech Republic\label{inst4}        
        \and Max-Planck-Institut f\"ur Radioastronomie, Auf dem H\"ugel 69, 53121 Bonn, Germany\label{inst5}     
     }

   \date{Received XXX; accepted XXX}

  \abstract
   {In addition to the supermassive black hole Sgr~A*, the inner parsec of our Galactic center is home to numerous fruitful scientific habitats. One of these environments is the S cluster, which consists of two distinct populations: the main-sequence S stars and the dusty G objects. While the majority of the brightest S stars can be classified as young B stars, the G sources can be described as dusty objects whose nature is still under debate.}
   {In this work, we focus on the most prominent G objects in the S cluster and follow their Keplerian trajectory around Sgr~A*. With this, we test the predictions based on almost two decades of monitoring of the direct vicinity of our central supermassive black hole using NACO and SINFONI, formerly mounted at the Very Large Telescope (VLT). The goal is to increase the existing data baseline for G2/DSO, D9, and X7 to get insights into their evolution on their Keplerian trajectories. In addition, we revisit the massive Young Stellar Object (YSO) X3 and scrutinize the potential impact of its environment on this highly dynamic source.}
   {The successor to the two instruments is called ERIS and offers upgraded optics and improved properties, including an enhanced spectral resolution. We utilize the IFU mode of ERIS, called SPIFFIER. We search for the Doppler-shifted Br$\gamma$ emission line to rediscover peculiar objects in the S cluster using SPIFFIER with the highest available spatial plate scale of 12.5 mas. Furthermore, we will derive the Br$\gamma$ luminosity of G2/DSO to inspect the degree of its change more than ten years after the pericenter passage. If present, a decrease in the Br$\gamma$ luminosity of G2/DSO on its descending part of the orbit would directly impact the direction of the debate about its nature.}
   {All the sources inspected in this work are rediscovered on their predicted astrometric positions, underlying the robustness of our monitoring efforts of the Galactic center. Furthermore, we find no signatures of a Br$\gamma$ luminosity variability for G2/DSO. Due to the enhanced capabilities offered by ERIS, we recovered the periodic pattern that resulted in the detection of D9, the first binary system in the S cluster. The bow shock source X7 does not show any deviation from its proposed Keplerian orbit and follows its projected orbit towards the north. Finally, we verify the presence of prominent stellar outflows for the massive YSO X3, consistent with the literature.}
   {}


   \maketitle
%

\section{Introduction} 
\label{sec:intro}

One significant advantage of the scientific habitat that envelopes the supermassive black hole Sgr~A* \citep[see][for recent reviews]{Eckart2017b,2022RvMP...94b0501G,2025arXiv250320081C} is the extensive data baseline that enables a detailed survey of individual sources. Based on the ongoing monitoring efforts of the Galactic center and, in particular, the direct vicinity of Sgr~A* using SHARP, the first Keplerian orbit of S2 was presented in \cite{Schoedel2002}. By including NACO K-band data, a first glimpse of the stellar content and the related Keplerian approximation of the S-cluster members was obtained by \cite{Gillessen2009}. Follow-up studies of the S-stars, which orbit Sgr~A*, showed an increased stellar density and pushed the number of detected cluster members by a factor of two \citep{Ali2020}. On even larger scales between 40 mpc and one pc, \cite{Fellenberg2022} and \cite{Jia2023} showed different, distinct stellar substructures following the disk detections by \cite{Paumard2006} and \cite{Lu2009}. Nowadays, it seems that the Nuclear Stellar Cluster partially consists of disk-like structures \citep{2009ApJ...697.1741B, Ali2020, Peissker2024a, Peissker2024b} that are dynamically interacting with each other \citep{Subr2009, Singhal2024}. This interaction can be visualized by distinctive orbital parameters, such as the inclination, the longitude of the ascending node, and the semi-major axis. The broad range of possible orbital configurations is analyzed in detail in \citet{Subr2009} and \citet{Haas2011}, where the authors include the gravitational interaction with the Circum-nuclear Disk that acts as a perturber on distances of a few parsecs with respect to Sgr~A*.

On smaller scales, \cite{Sukova2021} suggested that stellar transits might trigger quasiperiodic accretion events onto Sgr~A* as well as an intermittent increased outflow, which provides an interesting (but speculative) explanation for the observation of one of the brightest infrared flares of the supermassive black hole by \cite{Do2019}. In general, many attempts have been made to explain the radiative processes of Sgr~A* using a broad coverage of the electromagnetic spectrum \citep{Yusef-Zadeh2006, Eckart2008, Eckart2012}. Although the rapid variability of Sgr~A* was targeted mainly with a multiwavelength approach using different instruments and wavelengths \citep{Witzel2018, Witzel2021}, the superior sensitivity of the James Webb Space Telescope allowed for a direct observation of constant flickering of the supermassive black hole \citep{Yusef-Zadeh2025}.

These efforts are only possible due to the mentioned monitoring of the Galactic center. However, the heart of our Galaxy not only hosts a supermassive black hole and interesting stellar distributions, but it is also home to a peculiar class of objects: a population of faint dusty objects that can be found at different distances from Sgr~A* \citep{Gillessen2012,Eckart2013,Peissker2020b}. \citet{Zajacek2014} already suggested that these objects might be former binary systems that have migrated into the inner parsec and subsequently split up due to the Hills mechanism. \cite{stephan2016} followed up on this idea and proposed that the population of faint objects might be related to mergers driven by the von-Zeipel-Lidov-Kozai (vZLK) effect \citep{vonZeipel1910, Lidov1962, Kozai1962, Lithwick2011}. Regardless of the formation scenario (binary merger products or pre-main-sequence stars), the broad-band spectrum of the dusty objects is consistent with a rather compact (a few astronomical units), optically thick envelope (partially inflowing or outflowing) that is additionally asymmetric due to the bow-shock formation \citep{Zajacek2017}, which can address the reported near-infrared polarized emission \citep{Shahzamanian2016}. For a recent review of the characteristics and formation scenarios of dusty (G) objects, see \citet{2024arXiv241000304Z}.   

Recently, \cite{Peissker2024c} found evidence that the G-population \citep{Eckart2013, Peissker2020b} may indeed host pre- and post-mergers as proposed by \cite{Ciurlo2020}. The detection of the first binary system in the S cluster called D9 was possible because of the intense observation campaigns using SINFONI. Hence, the progress in understanding fundamental processes close to our supermassive black hole is only possible with an extensive data baseline that covers multiple wavelengths. These efforts also resulted in the observation of a putative accretion event of a thermal clump on the massive Young Stellar Object (YSO) system X3 \citep{Clenet2003a, muzic2010, peissker2023b}. The formation scenario of X3 is under debate and might be related to migration effects \citep{Bonnell2008, Hobbs2009, Jalali2014, Peissker2024b}.

In this work, we provide an update on the continued monitoring of dusty sources in the S cluster and focus on the most prominent objects, G2/DSO \citep{peissker2021c} and D9 \citep{Peissker2024c}, using the successor of the VLT instruments NACO and SINFONI, ERIS. Using the recently obtained ERIS data, we can close the gap in the data baseline between 2019 and 2024 to study peculiar objects in the S cluster.
In Section~\ref{sec:data}, we will list the used data. In Section~\ref{sec:results}, we will present the results of this observational campaign. Finally, the results will be discussed in Section~\ref{sec:discuss}, which is followed by a comprehensive conclusion in Section~\ref{sec:conclusion}.

\section{Data} 
\label{sec:data}

This section lists the data that was used and downloaded from the ESO archive\footnote{\url{https://archive.eso.org/eso/eso_archive_main.html}}. To provide an overview of the sources mentioned in Sec. \ref{sec:intro}, we present a finding chart in Fig. \ref{fig:finding_chart}. The data related to the finding chart will also be listed in Appendix A. {In contrast to previous studies \citep{Peissker2019, Peissker2020b, peissker2021c}, we are not applying the Lucy-Richardson algorithm to the data.}
\begin{figure}[htbp!]
	\centering
	\includegraphics[width=.5\textwidth]{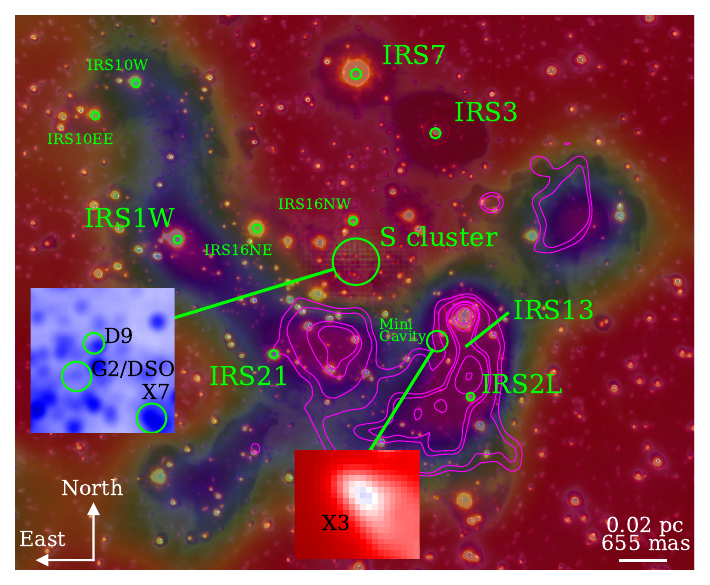}
	\caption{Multiwavelength finding chart of the inner $\approx$0.8 pc of the Galactic center. For this representation of the fraction of the inner parsec, we use ALMA continuum data observed at ($9.99\pm1.58$)$\times10^{10}$\,Hz, which is represented as green in the plot. The red background image and blue emission is observed with NACO in the K- and L-band, respectively. We overlay the plot with the magenta-colored Doppler-shifted Br$\gamma$ emission line observed at 2.1648\,\textmu m using ERIS and a related width of $\pm 330$\,km/s. The contour lines represent 9$\%$, 13$\%$, 26$\%$, 30$\%$, 43$\%$, and 65$\%$ of the peak emission of IRS 13 of about $\rm 0.9\times10^{-6}\,{\rm W/cm^2}$ \citep{Maillard2004}.}
\label{fig:finding_chart}
\end{figure}

\subsection{ERIS}

In 2022, the {first light} was announced for the Enhanced Resolution Imager and Spectrograph (ERIS) mounted at the Very Large Telescope (VLT). The system provides an imaging and Integral Field Unit (IFU) mode, which are called Near-Infrared Camera System (NIX) and Spectrometer for Infrared Faint Field Imaging Enhanced Resolution (SPIFFIER). ERIS-SPIFFIER is a direct upgrade to the Spectrograph for Integral Field Observations in the Near Infrared (SINFONI) and will be used for the data analysis in this work. Both systems, ERIS-NIX and ERIS-SPIFFIER, are fed with an adaptive optics system that allows the use of a natural guide star (NGS) and a laser guide star (LGS). In Table \ref{tab:eris_data}, we list all the data used for the analysis in this work. Since the standard deviation of the continuum is affected by strong absorption and emission lines in the spectrum, we use the continuum level to measure the quality of the Br$\gamma$ emission line. 
\begin{table*}[htb]
\centering
\caption{List of ERIS data used in this work.}
\begin{tabular}{ccccccccc}\hline \hline\\[-5pt]
Epoch  & ID & NDIT & DIT   &    PS  & S/N & FWHM$_x$ & FWHM$_y$ & Seeing\\
    &      &      & [sec]  &  [mas] &   & [mas]  & [mas] & [as]\\[3pt] \hline\\[-5pt]
2023.279    & 111.24H0.002     & 3  &   600    &  12.5      & 5.4 & 62.37$\pm$2.25 & 61.25$\pm$2.12 & 0.5-0.6 \\
2023.435    & 111.24H0.003     & 3  &   600    &  12.5      & 3.7 & 66.27$\pm$1.40 & 68.75$\pm$1.43 & 0.3-0.7 \\
2024.221    & 0112.B-2117(A) & 3  &   600    &  12.5        & 4.3 & 80.12$\pm$5.37 & 74.00$\pm$4.25 & 0.5-0.8 \\
2024.224    & 0113.B-0111(A) & 3  &   600    &  12.5        & 4.1 & 66.75$\pm$1.87 & 67.00$\pm$1.95 & 0.3-1.0 \\\hline
2023.443    &  111.24H0.003    & 14  &   30    &  100       & 5.8 & 107.10$\pm$4.45& 82.26$\pm$4.71 & 0.3-0.7 \\\hline
2023.443    &  111.24H0.003    &  7  &   10    &  250       & 6.1 & 333.86$\pm$6.81& 267.33$\pm$5.56 & 0.3-0.7 \\[3pt]
\hline \hline\\[-10pt]
\end{tabular}
\tablefoot{Please note that we measured the {Doppler shifted} Br$\gamma$ {line of G2/DSO} against the continuum emission to derive the S/N ratio {for the data with a spatial plate scale of 12.5 mas. For the data with a spatial plate scale of 100 and 250 mas, we use the bow shock source X3 because of crowding effects in the dense S cluster.} Hence, the indicated quantities describe a signal-to-continuum ratio and are a lower limit to describe the data quality. {We refer to Appendix C to outline the spectral analysis used to derive the S/N ratio. Furthermore, the FWHM of S2 indicates a qualitative parameter to compare the 12.5 mas data quality. For the 100 mas and 250 mas data, we list the FWHM of IRS 16NW and IRS 1W, respectively. The listed seeing in arcsec is taken from \url{www.eso.org} and indicates the \textit{DIMM-Seeing-at-start-time}.}}
\label{tab:eris_data}
\end{table*}
Hence, the S/N ratio indicated in Table \ref{tab:eris_data} can be considered as a lower level. In general, the data quality is superior compared to the SINFONI observations with comparable integration time \citep{peissker2021c, Davies2023}. For further information about ERIS, we refer the reader to \cite{Eris_1_2014, Eris_2_2014, Eris_3_2014} and \cite{Davies2018, Davies2023}.

\subsection{Additional data}

For Fig. \ref{fig:finding_chart}, we used ALMA and NACO data in addition to the ERIS observations presented in the former section. However, all the results presented in this work are continued monitoring efforts of the S cluster and a source in the mini-cavity close to IRS 13 that lasted for about two decades using mainly SINFONI and NACO  \citep{peissker2023c, Peissker2024b}. We refer the reader to \cite{peissker2021}, \cite{peissker2021c, peissker2023a}, and \cite{peissker2023b} for a detailed analysis of the sources.

\subsection{Data reduction}

The data reduction of the ERIS data used standard correction processes like FLAT fielding, DISTORTION correction, and WAVELENGTH calibration. We used GASGANO and ESOREFLEX to reduce the data \citep{Gasgano2012, Freudling2013}. The shown data are the stacked results of individual observations to increase the S/N level (Table \ref{tab:eris_data}). Remaining hot pixels were removed individually. Except for Fig. \ref{fig:finding_chart}, all images are homogenized with a 2-pixel Gaussian to increase the structures of the shown sources.

\subsection{Source selection}
In this work, we will focus on the kinematic and astrometric analysis of peculiar sources. The ongoing analysis of these sources directly addresses their proposed nature.

\subsubsection{G2/DSO}

As proposed in the literature, some authors are in favor of describing the source G2/DSO as a coreless gas cloud that orbits the SMBH in the S cluster \citep{Gillessen2012}. Although the scenario of a gas cloud that might partially be accreted by Sgr~A* is not supported by observations \citep{Witzel2014, Valencia-S.2015,Tsuboi2013, Tsuboi2015} and the comparisons of the data with the proposed models \citep{Shahzamanian2016,Shahzamanian2017,Zajacek2017,peissker2021c}, we will perform a check-up on the stability of G2/DSO in this post-pericenter phase.

\subsubsection{D9}

In \cite{Peissker2024c}, we present the first detection of a binary inside the S cluster (Fig. \ref{fig:finding_chart}). The binary system is among the G-object population and orbits Sgr~A* on a mildly eccentric trajectory. Given the vZLK timescales, it is possible that the binary will eventually merge \citep{Haas2016, Haas2021}. Following its orbit will reduce the uncertainties regarding the stellar parameters and might reveal expected fluctuations of its brightness.

\subsubsection{X7}

Another peculiar object in the S cluster is the bow-shock source X7, {first reported in \cite{muzic2007} and} analyzed in detail in \cite{muzic2010}, \cite{peissker2021}, and \cite{Shaqil2025}. The evolution of the elongation of X7 was shown in \cite{peissker2021}, where the authors proposed that a putative outflow from Sgr~A* might not be the origin of the bow-shock source. However, a definite answer for the origin of the bow-shock source is still under debate and requires precise knowledge of the orbit of X7. It should be noted that \cite{Ciurlo2023} {suggested} a comparable nature for X7 as it was claimed for G2/DSO {\citep{Gillessen2012}. Both Ciurlo et al. and Gillessen et al. claim, that X7 and G2/DSO are coreless dust features with a few {Earth masses} on a Keplerian trajectory orbiting Sgr~A*. Although a detailed discussion exceeds the scope of this work, we stress that} evaporation timescales {of about 10 years for dust with a few 100K and the ionized gas} of $\sim 10^4\,{\rm K}$ {challenge} the existence of {dense coreless} clouds in the S cluster \citep{Cuadra2005, Cuadra2006, Burkert2012, Peissker2024a}. {Therefore,} the proposed inspiral trajectory {for X7} will be inspected using the high-resolution ERIS data.

\subsubsection{X3}

The massive candidate YSO X3 is located about 0.1 pc away from the Sgr~A* and the S cluster. The object morphology reveals a prominent bow shock that is located inside the mini-cavity at a distance of less than 0.01 pc from the core region of IRS 13 \citep{Clenet2003a, Clenet2005a, muzic2010}, which is a peculiar dense association of early-type/Wolf-Rayet stars \citep{Pavlik2024,2025arXiv250421640L}. In \cite{peissker2023b}, it is shown that the K-band emission associated with the stellar source X3a is comoving with the envelope known as X3. Furthermore, it was possible to witness an active accretion event of a massive thermal blob onto X3a. We named this blob X3b and followed its trajectory for almost two decades. Intriguingly, X3b disappeared during the epoch when it came close to the tidal (Hill) radius of X3a. Hence, ongoing monitoring of this source allows us to confirm or reject the proposed interpretations. 

\section{Results}
\label{sec:results}

In this section, we will show the results of the observations and update existing orbital solutions for the dusty sources in the S cluster due to the increased data baseline. Using recently obtained data, we will close the observational gap between the SINFONI instrument (decommissioned in 2019) and ERIS \citep[first light in 2022, see][]{Davies2023}.
For X3, only observations with a medium plate scale are available (Table \ref{tab:eris_data}). Hence, we will focus on a quantitative analysis of the massive YSO.

\subsection{Source identification}
\label{sec:results_source_ident}

To increase the data baseline of peculiar sources in the Galactic center, it is crucial to identify the objects of interest in the observations executed with ERIS. Using the surveys and related Keplerian solutions presented in \cite{Peissker2020b, peissker2021, peissker2021c, peissker2023a, peissker2023b, Peissker2024a, Peissker2024c}, the ERIS data can be analyzed at the expected spatial position and the Doppler-shifted wavelength. {To estimate the related line-of-sight (LOS) velocity, we use a Gaussian fit of the Doppler-shifted Br$\gamma$ emission peak. With the known rest wavelength of the Br$\gamma$ line of $\rm 2.1661\mu m$, we calculate the LOS velocity of the source of interest.} 

\subsubsection{G2/DSO}

As for every other stellar source in the S cluster, the identification of G2/DSO suffered from confusion \citep{Sabha2012, Eckart2013} due to high crowding. Unfortunately, the confusion for G2/DSO was increased in epochs after its pericenter passage in 2014.43 due to the high stellar density close to Sgr~A* \citep{peissker2021c}. Based on the predicted orbit, this confusion decreases, especially in 2024. Using the high-resolution data listed in Sec. \ref{sec:data}, we identify G2/DSO in 2024.22 using ERIS at its expected position on a Keplerian orbit. We find no signatures of any deviation from its proposed trajectory as inferred by \cite{peissker2021c}.
\begin{figure}[htbp!]
	\centering
	\includegraphics[width=.5\textwidth]{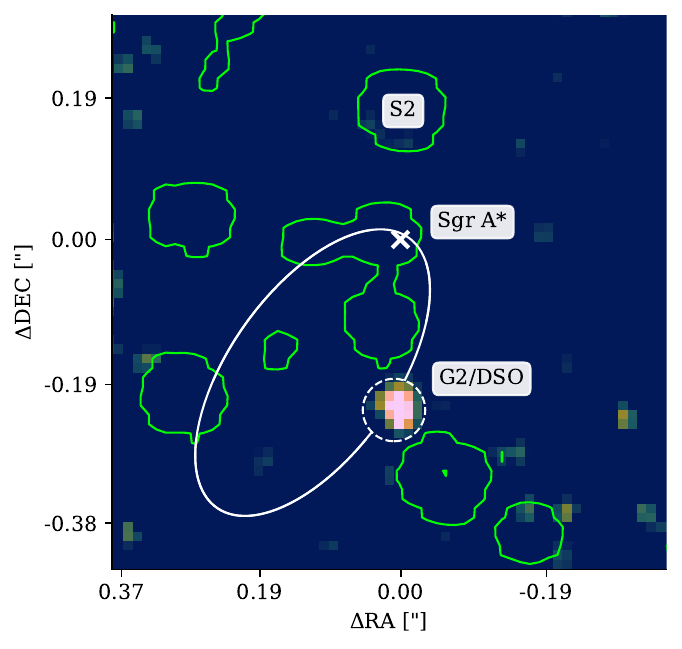}
	\caption{Detection of G2/DSO on its Keplerian orbit with ERIS in 2024. The Br$\gamma$ line map illustrates the preserved compact shape of G2/DSO and its continued path on a Keplerian orbit. The image is centered on Sgr~A* and shows continuum contour lines of the brightest S cluster stars. The location of S2, Sgr~A*, and G2/DSO is indicated. North is up, east is to the left.}
\label{fig:dso_ident}
\end{figure}
In Fig.~\ref{fig:dso_ident}, we show the identification of G2/DSO on its orbital path around Sgr~A*. 
With the detection in 2024, about 25$\%$ of the orbital period of about 100 years of G2/DSO is covered (Table \ref{tab:orbital_elements}). Another aspect of the analysis is the Br$\gamma$ luminosity of G2/DSO. Several predictions of a variable luminosity have been made \citep{Gillessen2012, Ballone2013, Pfuhl2015}. However, the data do not support significant variations as shown in \cite{Peissker2024a}. For the observational epoch in 2024, {we use the K-band flux estimate of $\rm 14\times10^{-3}$Jy from \cite{Peissker2024a} of S2 to normalize the IFU data. This flux estimate equals about 7$\rm L_{\odot}$ and can be used, to directly extract the Br$\gamma$ from the line map shown in Fig. \ref{fig:dso_ident}. We find} a continuation of the trend of a constant luminosity shown in Fig. \ref{fig:dso_mag}.
\begin{figure}[htbp!]
	\centering
	\includegraphics[width=.5\textwidth]{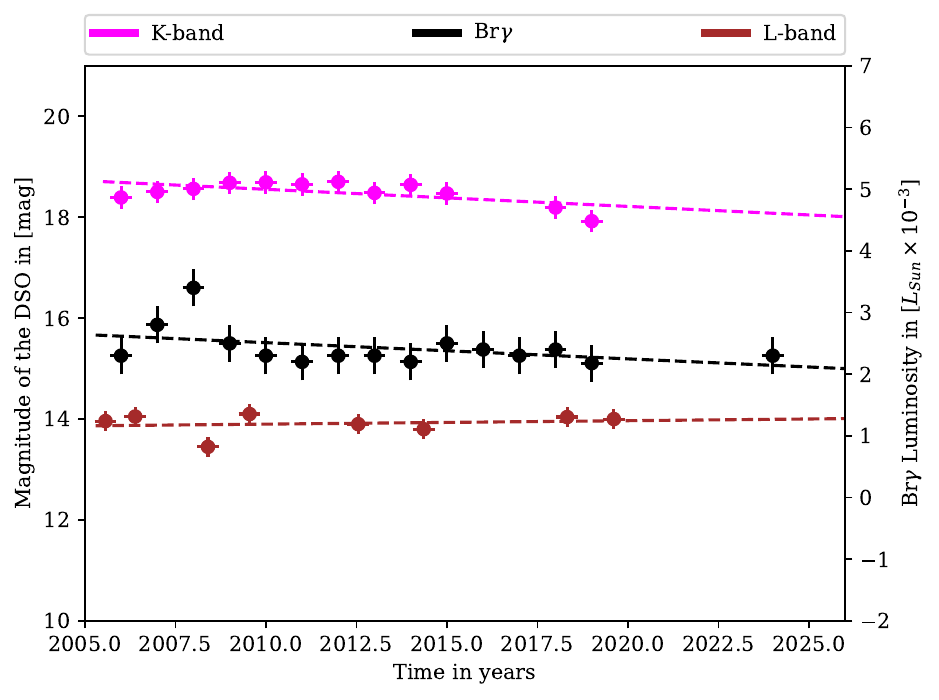}
	\caption{Magnitude and luminosity evolution of G2/DSO. The continuum detection is limited to 2019, whereas we added a new Br$\gamma$ luminosity data point representing the ERIS observations in 2024. As we mentioned in \cite{peissker2021c}, we find no signatures of variability for G2/DSO. Follow-up studies will focus on the continuum detection and evolution of the source.}
\label{fig:dso_mag}
\end{figure}
With an average Br$\gamma$ luminosity of $\rm L_{Br\gamma}\,=\,(2.42\pm0.30)\times10^{-3}L_{\odot}$, the estimated value is in agreement with \cite{Pfuhl2015} and \cite{Peissker2024a}. The measured luminosity in 2024 of G2/DSO is $\rm L_{Br\gamma}\,=\,(2.30\pm0.31)\times10^{-3}L_{\odot}$ where the uncertainty represents the standard deviation. Overall, we find no signs of any variability as expected from a stellar object that excites hydrogen at temperatures of several thousand Kelvin. 

\subsubsection{D9}

\begin{figure}[htbp!]
	\centering
	\includegraphics[width=.5\textwidth]{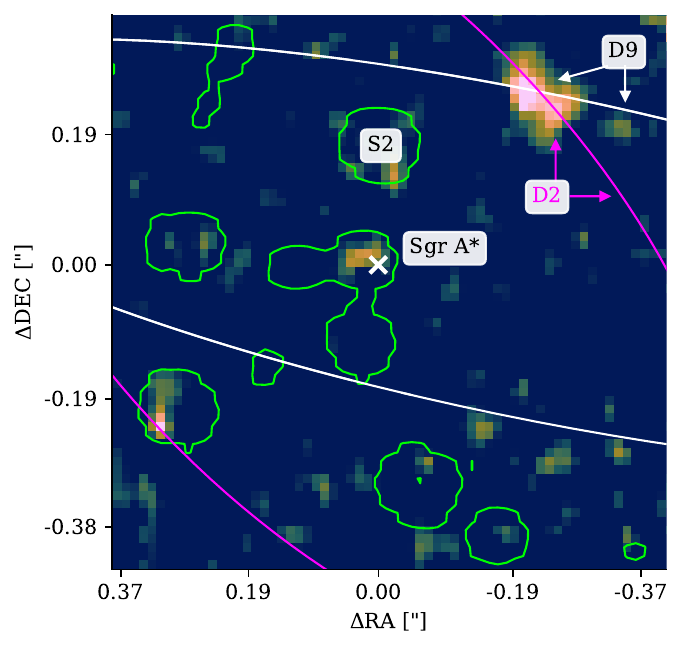}
	\caption{Doppler-shifted Br$\gamma$ line detection of D9 and D2 with ERIS in 2024. The white orbit plot indicates the trajectory of D9. Furthermore, the path and source are denoted with white arrows. In addition, we include the Keplerian approximation of D2 adapted from \cite{Peissker2020b}. The magenta arrows show the source and trajectory of the dusty object {D2} that was in superposition with D23 in 2019 \citep{Peissker2024a}. We expect increased confusion for both D9 and D2 from 2025 to 2027. {The dusty source D23 moved out of the FOV in 2024 and is not visible in this figure.}}
\label{fig:d9_ident}
\end{figure}

With the increased spectral resolution of ERIS, we identify the binary system D9 in 2023 and 2024 in its descending part of its orbit around Sgr~A*. As pointed out in \cite{Peissker2024c} and shown in \cite{Peissker2024a}, multiple sources are expected to cross the projected trajectory of D9. Hence, we test our predictions of the orbits of the dusty sources in 2024. As displayed in Fig. \ref{fig:d9_ident}, we identify D9 at the expected position on its Keplerian orbit around Sgr~A*. 
Furthermore, we find a source located about 25 mas away from D9. This source is moving towards north and was in superposition with D23 in 2019 \citep[see Fig. 1 in][]{Peissker2024c}. Based on the Keplerian approximations presented in \cite{Peissker2020b} and \cite{Peissker2024a}, we successfully rediscover D2 on its orbit\footnote{Note that \cite{Sitarski2016} and \cite{Ciurlo2020} refer to D2 as G3.}. It is expected that both sources will be superposed between 2025 and 2027. Planned observations should take this potential source of confusion into account.

\begin{figure}[htbp!]
	\centering
	\includegraphics[width=.5\textwidth]{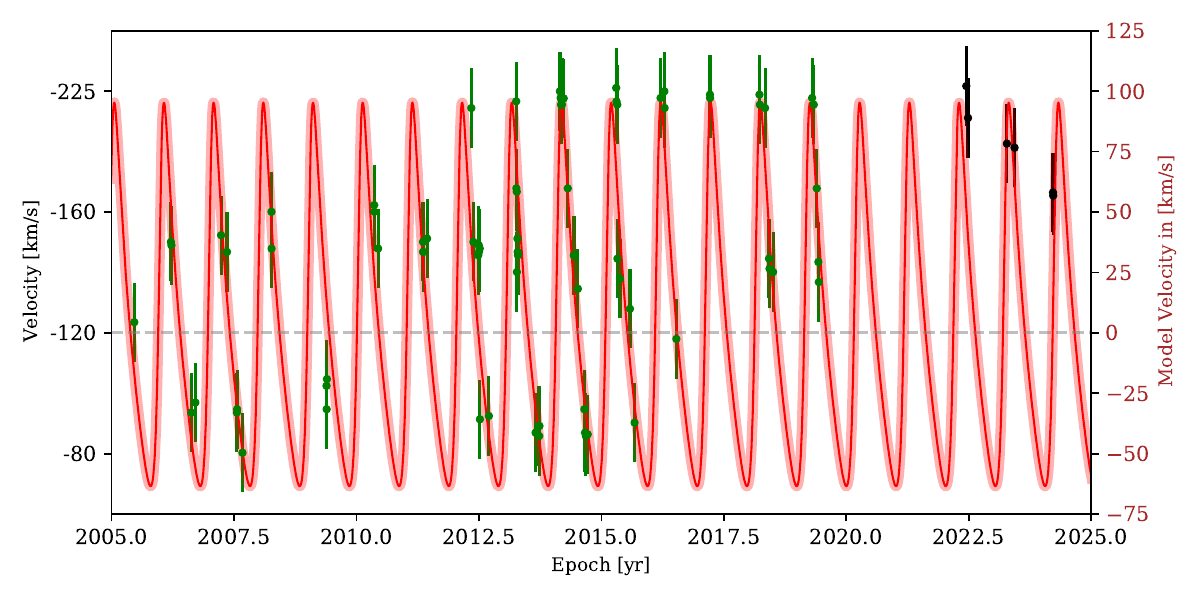}
	\caption{Resulting periodic plot of the D9 binary system. The plot does cover almost 20 years of ongoing IFU efforts in the Galactic center, including SINFONI and ERIS. The black data points are based on ERIS observations and are at their expected spectroscopic position. For the gap between 2019 and 2022, no IFU observations of the Galactic center are available.}
\label{fig:d9_periodic}
\end{figure}

Considering the Doppler-shifted and periodically varying Br$\gamma$ emission line of the binary system, we apply corrections listed in Table \ref{tab:d9_corrections} to the observations of D9 in 2023 and 2024.
\begin{table}[htb]
\centering
\caption{Data points of the binary system D9 observed in 2023 and 2024.}
\begin{tabular}{ccccc}\hline \hline
Epoch    &    Line  &  Barycenter  &  Base  &   Velocity  \\
         & [\textmu m] & [km/s] & [km/s] & [km/s] \\\hline 
2023.279 & 2.16452  & 16.14 & 124.41  & 78.28  \\ 
2023.435 & 2.16463  & 2.59  & 124.41  & 76.60  \\
2024.221 & 2.16465  & 18.33 & 124.41  & 58.09  \\
2024.224 & 2.16466  & 18.27 & 124.41  & 56.76  \\
\hline \hline
\end{tabular}
\tablefoot{The detected Br$\gamma$ emission and normalized baseline are indicated. Furthermore, we list the barycentric correction and the resulting final velocity. Please see \cite{Peissker2024c} for further information about the correction.}
\label{tab:d9_corrections}
\end{table}
The corrections listed in Table \ref{tab:d9_corrections} include normalization on the baseline and include the intrinsic arrangement of the binary, both incorporated in column {Base}. The barycentric correction is abbreviated with {Barycenter} and applied in the final velocity. The resulting velocities are shown in Fig. \ref{fig:d9_periodic}, which extends the baseline by two epochs compared to \cite{Peissker2024c}.
For the inspected Doppler-shifted Br$\gamma$-line, we find the periodic signal at its predicted spectroscopic position. {With a spectral resolution of ERIS in the inspected K-band range of $R=10000$, the $1\sigma$ uncertainty is about 15 km/s and comparable to the SINFONI observations. Accounting for systematics, common spectral uncertainties are in the range of $10-15$ km/s \citep{gravity2018}. As pointed out by \cite{Peissker2024c}, the normalization to the data baseline of $\rm \approx\, 125$ km/s for D9 is fitted for the epochs between 2013 and 2019. This means that the data baseline needs to be updated in the near future if more data is available. For the inspected ERIS data of 2024, we apply a conservative uncertainty of $\rm \pm\,15$ km/s to each data point to cover the variation of the LOS velocity of D9 on the descending part of its orbit and instrumental/observational systematics.} Follow-up studies need to address this change appropriately.

\subsubsection{X7}

\begin{figure}[htbp!]
	\centering
	\includegraphics[width=.5\textwidth]{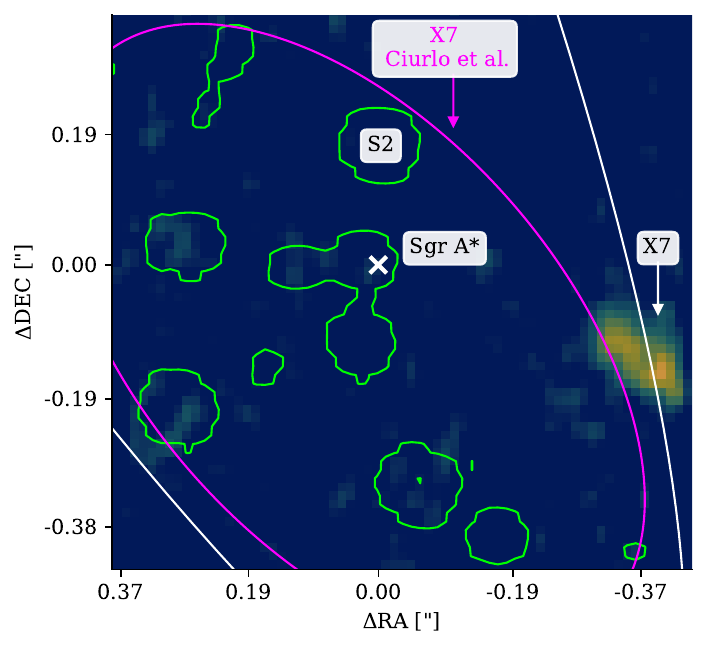}
	\caption{Detection of X7 on its designated Keplerian orbit around Sgr~A*. The magenta-colored orbit depicts the Keplerian solution presented in \cite{Ciurlo2023}, whereas the white trajectory is adapted from \cite{Peissker2024a}. While the center of gravity of the bow shock source is located on the white orbit, the tip of X7 is following the magenta trajectory. {We want to stress that the limited FOV blocks emission of X7. In this representation, only about 1/3 of the bow shock is visible \citep{peissker2021, Ciurlo2023}}.}
\label{fig:x7_ident}
\end{figure}

Another peculiar object of the S cluster is the bow shock source X7 \citep{Clenet2003a, muzic2010, peissker2021, Ciurlo2023}. Until now, different scenarios for X7 have been discussed, reflecting the prominent elongation first reported in \cite{peissker2021}. In this work, we will not discuss an interpretation of the data, but rather the astrometric detection of X7 itself. Because of the limited Field of View (FOV) of the ERIS and SINFONI IFU, a comprehensive inspection of the Doppler-shifted Br$\gamma$ line is limited \citep{peissker2021}. Hence, the Keplerian elements presented in this work have a tendency to display the leading part of X7\footnote{Please see Figure 4 in \cite{peissker2021}.}. Compared to \cite{peissker2021}, we will only use the Br$\gamma$ emission to fit a Keplerian orbit to the trajectory of X7 because of the missing L band observations. In Fig. \ref{fig:x7_ident}, we show the observation of X7 in 2024 with its prominent elongation.
Considering a projected length of the dust emission of X7 of about 0.3 arcsec in 2021 as displayed in \cite{Ciurlo2023}, our measurements of the Doppler-shifted Br$\gamma$ line emission only covers about 33$\%$ due to the limited FOV (Fig. \ref{fig:x7_ident}). However, we find a satisfying agreement with our Keplerian approximation as shown in Fig. \ref{fig:x7_ident}. Due to the orientation of X7 and its proper motion, that is, directed towards the north, we find no indications that the head significantly approaches Sgr~A*. However, this does not entirely exclude that some material might evaporate or gets redirected by leaving a bound Keplerian orbit.

\subsubsection{X3}

\begin{figure*}[htbp]
	\centering
	\includegraphics[width=1.\textwidth]{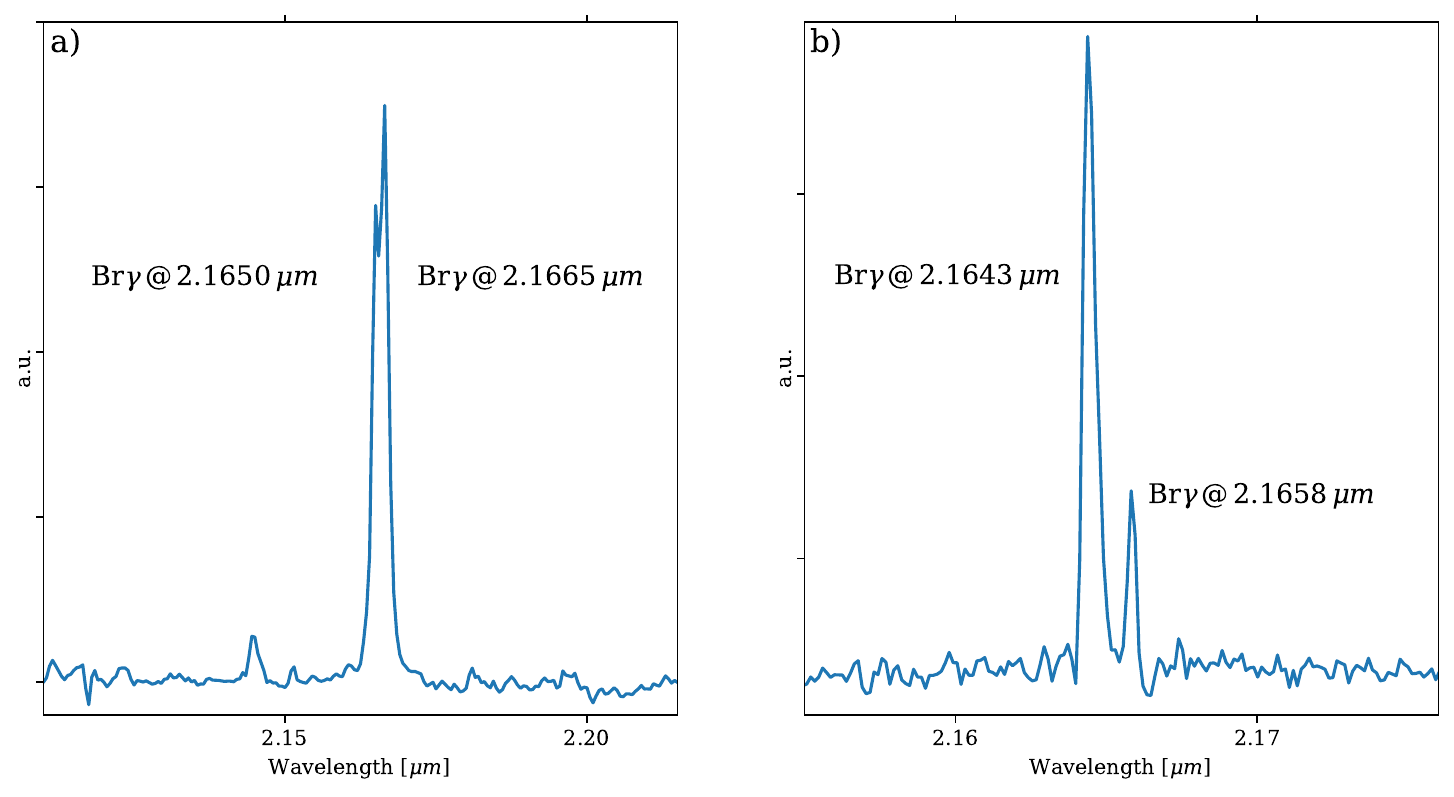}
	\caption{Br$\gamma$ emission doublet of X3 observed with {SINFONI (left plot, a)) and ERIS (right plot, b)) with a plate scale of 100 mas in 2014 and 2023, respectively. For both Doppler-shifted Br$\gamma$ peaks in 2014 and 2023, we determine a separation of $\rm 1.5\times10^{-3}\mu m$ that equals 200 km/s \citep{peissker2023b}. The peak values of the lines are listed in Table \ref{tab:X3_line_evo}.}}
\label{fig:x3_spec}
\end{figure*}

The majority of sources investigated in this work are located in the S cluster. However, one intriguing object can be found in the mini cavity close to IRS 13 (Fig. \ref{fig:finding_chart}). This source is classified as a massive YSO with a prominent bow shock and is called X3 \citep{Clenet2003a, muzic2010, peissker2023b}. Because it is located outside of the S cluster with an approximately low mean proper motion of about 240 km/s, we focus on a spectroscopic analysis to verify some of the observed features, such as the double {peak} of the Br$\gamma$ emission line \citep{peissker2023b}. Due to the lack of observations with the highest plate-scale, we use the 100 mas setting of the ERIS observation executed in 2023. The resulting spatial pixel scale is 50 mas. In Fig. \ref{fig:x3_spec}, we show the Br$\gamma$ emission line of X3. 
Compared to the spectrum of X3 observed in 2014 with SINFONI, we notice an increased intrinsic system velocity for the rediscovered double-peaked Br$\gamma$ emission line. In order to determine the system LOS velocity, we take the average of Br${\gamma}_{\rm blue}$ and Br${\gamma}_{\rm red}$. With the velocities listed in Table \ref{tab:X3_line_evo}, we determine a blueshifted system velocity of X3 of $-48.5$ km/s in 2014. The LOS velocity for the X3 bow shock system increased to $-145.4$ km/s in 2023.

\begin{table}[htbp!]
\centering
\caption{Comparison of the two Br$\gamma$ doublet detections with SINFONI and ERIS in 2014 and 2023, respectively.}
\begin{tabular}{ccccccc}\hline \hline\\[-5pt]
Year  &  Br${\gamma}_{\rm blue}$  & LOS$_{\rm blue}$ & Br${\gamma}_{\rm red}$ & LOS$_{\rm red}$ & $\Delta\nu$ & $\Delta \rm v$ \\ 
      &        [\textmu m]        & [km/s] &     [\textmu m]        & [km/s] & [\textmu m] & km/s           \\[3pt] \hline\\[-5pt]
2014 & 2.1650  & -152.34 & 2.1665 & 55.39 & 0.0015 & 207.73 \\ 
2023 & 2.1643  & -249.29 & 2.1658 & -41.54& 0.0015 & 207.75 \\[3pt]
\hline \hline
\end{tabular}
\tablefoot{For both epochs, the doublet is 207 km/s apart, suggesting that the origin of both emission lines is associated with X3. {We list the related LOS velocity of the measured Doppler-shifted peak wavelength.} As reported, the Br$\gamma$ emission lines might be related to outflows of the massive YSO \citep{peissker2023b}.}
\label{tab:X3_line_evo}
\end{table}
In addition to the Br$\gamma$ doublet shown in Fig. \ref{fig:x3_spec}, we find some weak signatures of an inverse P Cygni profile depicted in the same figure. 
The weak signal is spectroscopically located at 2.1667\,\textmu m, which corresponds to a red-shifted velocity of 83 km/s. The lower Br$\gamma$ emission intensity of the line at 2.1658\,\textmu m  might be influenced by the inverse P Cygni absorption feature \citep{Carr2022}. {Due to the noise, especially close to the Br$\gamma$ rest wavelength at $2.1661\mu m$ and the weakness of this putative spectral feature, an artificial origin cannot be excluded.}

\subsection{Keplerian orbits}

To verify or update the orbital solutions of G2/DSO, D9, and X7, we use the ERIS observations executed in 2023 and 2024. {Based on the established reference frame using the SiO masers \citep{Menten1997, Plewa2015}, the orbit of S2 and hence, the position of SgrA*, is known using different telescopes and instruments \citep{Parsa2017, gravity2018, Do2019S2}. With a precision of 10$\%$ of a pixel with a spatial pixel scale of 12.5 mas, the position of Sgr~A* is well known. Including the ongoing monitoring efforts with GRAVITY, the uncertainties for the position of Sgr~A* are in the range of a few $\rm \upmu as$.}\newline
From the {spectral} identification of the Doppler-shifted Br$\gamma$ line {and astrometric detection of the sources of interest} presented in Sec. \ref{sec:results_source_ident}, we expand the existing data baseline from 2005 to 2024. The results of the analysis are already incorporated as orbits in the plots (Fig. \ref{fig:dso_ident}, \ref{fig:d9_ident}, and \ref{fig:x7_ident}). {Using a distance of 8 kpc and a mass for the central gravitational potential Sgr~A* of $\rm 4\times10^{6}M_{\odot}$ \citep{eht2022, Peissker2022}, the derived} Keplerian elements for the S cluster sources G2/DSO, D9, and X7 are listed in Table \ref{tab:orbital_elements}.
\begin{table*}[htbp!]
    \centering
    \caption{Keplerian elements of the S cluster objects analyzed in this work using ERIS.}
    \setlength{\tabcolsep}{0.999pt}
    \begin{tabular}{|cccccccc|}
            \hline
            \hline
            Source & $a$    & $e$ & $i$        & $\omega$   & $\Omega$   & $t_{\rm closest}$ & T$_P$   \\
                   &  [mpc] &     & [$^\circ$] & [$^\circ$] & [$^\circ$] &  [years]          & [years] \\

            \hline
             {G2/DSO} - old &17.23 $\pm$  0.20   &0.963 $\pm$ 0.004  &120.32 $\pm$ 2.40   & 92.81 $\pm$  1.60  &  63.02 $\pm$ 1.37   & 2014.43 $\pm$ 0.01   & 105.90\\
             {G2/DSO} - new &17.25 $\pm$  0.10   &0.965 $\pm$ 0.002  &121.03 $\pm$ 1.71   & 93.85 $\pm$  2.01  &  63.59 $\pm$ 1.94   & 2014.43 $\pm$ 0.01   & 106.10\\
             \hline
             D9 - old & 44.00 $\pm$ 1.145  &  0.32 $\pm$ 0.01  & 102.55 $\pm$ 1.14  & 127.19 $\pm$ 8.02  & 257.25 $\pm$ 1.71   & 2309.13 $\pm$ 7.01   & 432.23\\
             D9 - new & 43.69 $\pm$ 0.040  &  0.31 $\pm$ 0.01  & 103.13 $\pm$ 0.45  & 136.36 $\pm$ 4.18  & 257.54 $\pm$ 4.75   & 2321.70 $\pm$ 0.01   & 427.67\\
            \hline
             X7 - old & 149.99 $\pm$ 0.36  & 0.76 $\pm$ 0.05   & 74.82 $\pm$ 7.84   & 190.22 $\pm$ 17.47 &  25.15 $\pm$ 10.77  & 1975.85 $\pm$ 0.30   & 2720.42\\ 
             X7 - new & 149.98 $\pm$ 0.18  & 0.76 $\pm$ 0.03   & 74.48 $\pm$ 6.87   & 189.07 $\pm$ 10.88 &  25.21 $\pm$ 12.60  & 1975.86 $\pm$ 0.01   & 2720.15\\
             \hline
    \end{tabular}
    \tablefoot{The row depicted with {old} is related to the data baseline that ends with the decommissioned instrument SINFONI in 2019 {and refers to the published orbital solutions in \cite{Peissker2024a}.} The Keplerian elements that are indicated with {new} include the ERIS data that was obtained until 2024. {The uncertainties for the new estimated Keplerian approximations are adapted from the MCMC simulations shown in the Appendix.}}
    \label{tab:orbital_elements}
\end{table*}
For all three sources listed in Table~\ref{tab:orbital_elements}, we only find marginal differences between the published orbital elements and this work. {Considering the uncertainties of the updated Keplerian elements which are adapted from the MCMC simulations shown in the Appendix, we find marginal differences. While some parameters seem to be constrained in a tighter uncertainty range, such as the semi-major axis, others exceed existing values (e.g., the longitude the ascending node). We interpret the relation between the old and new uncertainties as a reflection of the noise level of the data. Inspecting the average values of the uncertainty range listed in Table \ref{tab:orbital_elements}}, we find deviations between the published orbital elements and this work of about $0{-}2\,\%$. Qualitatively, this is reflected by the confusion-free detection of the sources as shown in Fig. \ref{fig:dso_ident}, \ref{fig:d9_ident}, and \ref{fig:x7_ident}. Since the Keplerian elements provided in \cite{Peissker2024a} are based on a data baseline that covers the epochs 2005 to 2019, imprecise orbital solutions would be reflected by astrometric deviations of the investigated source sample. However, comparing both orbital solutions, we find an averaged deviation of the six Keplerian elements between the old and new orbit of $0.48\,\%$ for G2/DSO, $1.63\,\%$ for D9, and $0.21\,\%$ for X7. The small deviations are expected since we find all sources at their projected astrometric position as we show in Fig. \ref{fig:dso_ident}, Fig. \ref{fig:d9_ident}, and Fig. \ref{fig:x7_ident}. {However, the absolute uncertainties of the Keplerian elements are only marginally affected by the longer data baseline which is due to the low orbital coverage of the observations \citep{ONeil2019, Ali2020}.} The full Keplerian approximations, including the astrometric measurements, are shown in Appendix B.

\section{Discussion} 
\label{sec:discuss}
In this section, we will discuss our findings using NIR ERIS observations of the Galactic center using the Doppler-shifted Br$\gamma$ emission line. The IFU data observed until 2024 reveals an unprecedented precision of the Keplerian analysis of peculiar objects in the S cluster. {We will present a qualitative perspective on the SINFONI and ERIS data to underline the capabilities of the new instrument mounted at the VLT.} Furthermore, the Doppler-shifted Br$\gamma$ doublet of X3 can be confirmed.

\subsection{A comparison of the SINFONI and ERIS data}

\begin{figure}[htbp]
	\centering
	\includegraphics[width=.5\textwidth]{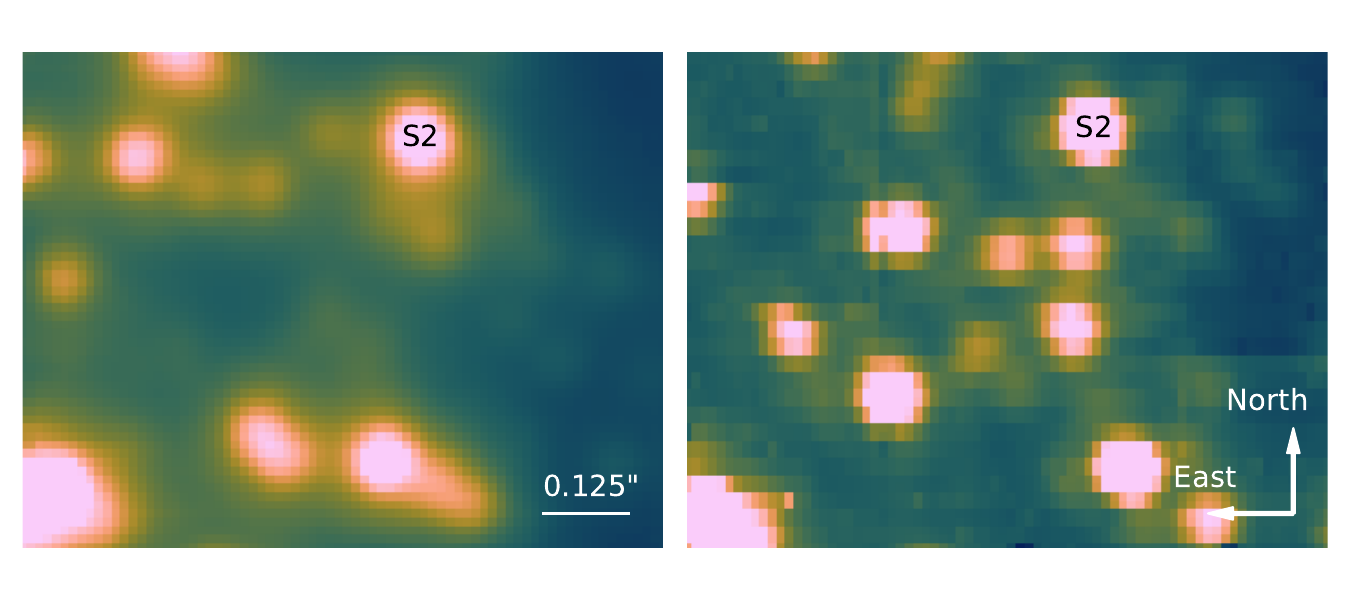}
	\caption{Comparison of S cluster observations using SINFONI (left) in 2014 and ERIS (right) in 2024. While the left image is the final product of about 50 hours of on-source observations, the right plot was obtained with an integration time of 30 minutes (including overheads). Due to PSF distortions, the left image seems to be smeared. Artifacts in the right image are the result of the insufficient mosaic recipe of ESOREFLEX and the low number of on-source integration time (Table \ref{tab:data_quality}). Both images have a spatial pixel scale of 12.5 mas and show the B2V star S2.}
\label{fig:sinfo_eris_comp}
\end{figure}

Given the monitoring efforts using SINFONI in the Galactic Center by targeting the innermost stars that orbit Sgr~A* between 2005 and 2019, we picked three example epochs to compare them with the recent ERIS observations. Because of the dense stellar crowding of the S cluster that is reflected by confused stars or varying background emission that may impact the apparent magnitude, we focus on the brightest S cluster member in the K band, namely S2 \citep{Habibi2017}. A qualitative strategy is to compare the FHWM of S2 by fitting a Gaussian to the K band data to measure the impact. Of course, different weather conditions or the general instrument performance could alter the FWHM. Since these sources of influence can be accounted for ERIS and SINFONI, Table \ref{tab:data_quality} provides a snapshot of the data quality of both IFU instruments and the stability of the FWHM.
\begin{table}[htb]
\centering
\caption{Qualitative comparison between SINFONI and ERIS data using the PSF of the brightest S cluster member S2 in the K-band.}
\begin{tabular}{ccccc}\hline \hline
Epoch & Integration & FHWM$_x$ & FWHM$_y$ & Area$_{\rm eff}$ $\times\,10^3$  \\
      &  [hours]     &[mas] & [mas] & [mas$^2$]  \\ \hline 
      & \multicolumn{3}{c}{SINFONI} & \\      
\hline     
2008.2 &   3.5   &56.6$\pm$1.7&53.2$\pm$1.5&2.3$\pm$0.1\\
2014.3 &  24.2   &72.3$\pm$2.3&76.2$\pm$2.5&4.3$\pm$0.2\\
2018.4 &  19.0   &70.6$\pm$1.7&76.0$\pm$1.6&4.2$\pm$0.1\\
\hline
& \multicolumn{3}{c}{ERIS} & \\      
\hline 
2023.2  &0.5       &59.8$\pm$4.5&61.7$\pm$4.5&2.8$\pm$0.3\\
2023.4  &0.5       &65.6$\pm$3.6&62.8$\pm$3.1&3.2$\pm$0.2\\
2024.2  &0.5       &68.6$\pm$6.5&68.7$\pm$4.7&3.6$\pm$0.4\\
\hline\hline
\end{tabular}
 \tablefoot{Here, we compare the effective PSF of SINFONI and ERIS to demonstrate the stability of the PSF in combination with the resulting quality after stacking individual data cubes to a resulting mosaic. Please compare the listed values here with the independent measured quantities in Table \ref{tab:data}. As is evident, the yield from short on-source integration observations using ERIS is high.}
\label{tab:data_quality}
\end{table}
As it is evident from Table \ref{tab:data_quality}, the PSF of ERIS is more stable despite a lower on-source integration time. The investigated improvements are consistent with the first-light analysis by \cite{Davies2023} and underline the increase in performance of ERIS over SINFONI by almost 10$\%$ in the K band with a nominal spatial pixel scale of 25 mas as it is investigated in \cite{George2016}.\newline
The above discussed improvements are reflected in an increased image quality that is demonstrated in Fig. \ref{fig:sinfo_eris_comp}.
While the ERIS data mosaic exhibits some imperfections, the major difference between the two images is the integration time. The SINFONI data on the left observed in 2014 in Fig. \ref{fig:sinfo_eris_comp} is a culmination of 50 hours of on-source integration time. In contrast, the ERIS observation is constructed from three individual points with 600 sec each, resulting in a total integration time of about 0.5 hours. Although the difference of the integration time in Fig. \ref{fig:sinfo_eris_comp} is in the order of two magnitudes, the ERIS image seems sharper. This is not a coincidence but due to the slightly decreased FWHM of the ERIS data \citep{George2016, Davies2023}. The reason is the improved optics of ERIS, which effectively reduces PSF distortions.\\
{Considering the spectroscopic measurements of both SINFONI and ERIS, statistical and systematic uncertainties affect the detection of the Doppler-shifted Br$\gamma$ line. We have shown that the OH line transitions (9-7) exceed statistical and systematic uncertainties by a factor of 10-30 \citep{peissker2021c}. Statistical uncertainties resulting from, for example, a low airmass, high seeing, decreased atmospheric turbulence, and detector performance are challenging to determine. A compromise between these sources of influence are the 1$\sigma$ uncertainties mentioned in Sec. \ref{sec:results}. Especially the Keplerian velocity approximation of the well observed B2V star S2 shows that spectroscopic uncertainties of $\pm 15$km/s for the here discussed IFU data are justified \citep{gravity2018}.}

\subsection{Astrometric precision}

Based on the Keplerian approximation presented in \cite{Peissker2020b} and \cite{Peissker2024a}, we have traced the most prominent dusty sources in the S cluster at their expected astrometric positions in 2024. With these additional positions, we derived the Keplerian orbital solutions listed in Table \ref{tab:orbital_elements} and classified them as new (i.e., this work) and as old \citep[i.e.,][]{Peissker2020b, peissker2023a, Peissker2024a}. We compare every individual old and new orbital element and derive an individual deviation listed in Table \ref{tab:deviations_orbital_elements}.
\begin{table}[htbp!]
    \centering
    \caption{Deviations (in \%) between existing and newly derived orbital elements for the most prominent dusty objects of the S cluster.}
    \begin{tabular}{|ccccccc|}
            \hline
            \hline
            Source & $\Delta a$    & $\Delta e$ & $\Delta i$        & $\Delta \omega$   & $\Delta \Omega$   & $\Delta t_{\rm closest}$  \\
            \hline
             {G2/DSO} & 0.11 & 0.20 & 0.59  & 1.12  & 0.90 & 0.00 \\
             D9  & 0.71 & 0.66 & 0.56 & 7.20 & 0.11 & 0.54 \\
             X7  & 0.01 & 0.00 & 0.46 & 0.61 & 0.23 & 0.00 \\ 
            \hline
    \end{tabular}
    \tablefoot{We use the Keplerian elements based on observations until 2019 (classified as {old} in Table \ref{tab:orbital_elements}) as a reference and calculate the percentage of the deviations compared to the updated values. {Please note that we used the averaged individual Keplerian elements of the indicated uncertainty range. The absolute uncertainties are estimated using MCMC simulations (see the Appendix) and exceed the values listed here.}}
    \label{tab:deviations_orbital_elements}
\end{table}
As stated in Sec. \ref{sec:results}, the averaged uncertainty for G2/DSO, D9, and X7 is $0.48\,\%$, $1.63\,\%$, and $0.21\,\%$, respectively. For the investigated source sample, we estimate an accumulated and averaged uncertainty of $0.77\,\%$, which suggests precise knowledge about the previous, current, and upcoming orbital evolution. However, for all three sources, secular gravitational interactions will induce a non-Keplerian evolution that we will discuss in the following subsection. {Furthermore, we want to stress that the nature and orbital evolution of X7 is still under debate \citep{peissker2021, Ciurlo2014, Shaqil2025}. We can not exclude the possibility that the orbital elements of X7 will change dramatically in the next observation epochs.}

\subsection{The origin of G2/DSO}

The most prominent dusty source of the S cluster is G2/DSO. As the name suggests, an ongoing debate about its nature results in a different morphology and fate of the dusty source. We will redirect the interested reader to \cite{Zajacek2014} and \cite{peissker2021c} for a detailed discussion of different scenarios. Among them, an inspiral trajectory of G2/DSO as part of a gas streamer is proposed \citep{Pfuhl2015}. This picture was updated in \cite{Gillessen2019} by claiming the presence of a drag force that acts on the gas. The authors of Gillessen et al. acknowledge the potential presence of a faint source and describe an embedded stellar system enveloped by gas and dust. To resolve the contrary conclusion about the nature of G2/DSO, the authors of \cite{Gillessen2019} propose a non-Keplerian motion of the gas that gets unavoidably disrupted by the central faint stellar source. According to Gillessen et al., the deviation of the gas (i.e., Doppler-shifted Br$\gamma$ emission) from a Keplerian orbit should be 5 mas in 2019 and 10 mas in 2021. In 2024.22, we rediscover G2/DSO at a Gaussian fitted distance to Sgr~A* of $\rm \Delta R.A.\,=\,0.021$ arcsec and $\rm \Delta DEC\,=\,0.22$ arcsec. Compared to the Keplerian orbit, we find a deviation of 5 mas to the measured astrometric position of G2/DSO. Considering the spatial pixel scale of the ERIS data of 12.5 mas, the estimated deviation of 5 mas is only a fraction of a pixel. Although this component scenario could solve the discrepancy between different proposed natures for G2/DSO, the uncertainty does not allow a definite statement. We point out that GRAVITY \citep{gravity} and the recent upgrade GRAVITY+ \citep{Gravity_plus_Collaboration2022} could reduce the astrometric uncertainty by several magnitudes. Given that the Br$\gamma$ luminosity of G2/DSO is $\rm L_{Br\gamma}\,=\,(2.30\pm0.31)\times10^{-3}\,L_{\odot}$ in 2024, the chance of observing the source with GRAVITY+ is plausible.
{Considering the constant continuum magnitude, Br$\gamma$ luminosity, and astrometric position, we can not confirm signatures of an in-spiral motion of G2/DSO as suggested by \cite{Gillessen2019, Gillessen2025}\footnote{{We note that the recently found object G3 by \cite{Gillessen2025} shares astrometric and spectroscopic similarities with OS2 \citep{peissker2021c}.}}.}

\subsection{The evolution of D9}

The vZLK timescale of $\tau_{\rm vZLK}^{\rm SMBH}\,=\,1.1\times 10^6$ years and the related upper age estimate of the binary system of about $2.7\times 10^6$ years are similar \citep{Peissker2024c}. Hence, the primary (D9a) and secondary (D9b) might merge soon. Thus, a precise monitoring of the orbit of the system is crucial for the detection of any deviations in the Keplerian motion that would indicate the approach of the merger of D9a and D9b \citep{Huang1956, Hadjidemetriou1963, Stone2013}. Given the deviation between the orbital elements for the data baseline that covers observations until 2019 and 2024 of 1.63$\%$, we find no signatures that could indicate an upcoming merger. However, the periodic signal of the binary does follow the fit presented {in} \cite{Peissker2024c}. We note that the upcoming superposition with D2 \citep{Eckart2013, Peissker2020b, Peissker2024a} will have an impact on the detection of D9 because of increased confusion.

\subsection{X7 -- a tidally stretched gas streamer?}

\cite{Ciurlo2023} proposed the interesting idea that X7 is a tidally stretched gas filament. We have {discussed this idea in the Appendix of \cite{peissker2021} and provided a data interpretation} in \cite{Peissker2024a}. {In this work, we focus} on the astrometric detection and the related implications. Consistent with the orbit proposed in \cite{Peissker2024a} that covers a data baseline of about 20 years, we rediscover X7 at its expected position in 2023 and 2024 using ERIS. Due to the missing L-band coverage {in} the here analyzed ERIS data, we limit the analysis to the Doppler-shifted Br$\gamma$ emission line as shown in Fig. \ref{fig:x7_ident}. In both epochs observed with ERIS, we fit a Gaussian to the tip region of the bow shock, which introduces some limitations. In other words, because the FOV does not cover the complete source, our projected data points suffer from a systematic offset towards Sgr~A*. However, we find an agreement between both Keplerian approximations using data until 2019 and 2024 of 0.21$\%$. This suggests that X7 is still following our proposed orbital trajectory and does not show any tidally stretched emission along its orbit as it is modeled for comparable sources under the impact of the supermassive black hole \citep{Schartmann2012, Ballone2013, Gillessen2013a}. Given the prominent elongation of X7, a mass transfer between X7 and Sgr~A* cannot be fully excluded. As it is obvious from Fig.~\ref{fig:x7_ident}, the authors of \cite{Ciurlo2023} use the tip of X7 {itself} to fit a Keplerian orbit\footnote{{Please see Fig. 11 of \cite{Ciurlo2023} to inspect the methods used for fitting an orbit to the trajectory of X7.}}. If the gravitational pull of Sgr~A* can explain the elongation of X7, we expect a detachment of the tip in the upcoming years due to the ongoing elongation.

\subsection{The origin of the massive YSO X3}

In addition to the formation scenarios proposed in \cite{Jalali2014} and \cite{peissker2023b}, the acceleration of about $\Delta v\,\approx\,100$ km/s of X3 suggests that the system could be bound to a gravitational potential. If we assume that X3 has formed outside the inner parsec, we would expect a significant deceleration as proposed in \cite{Hobbs2009}. But if X3 had formed due to some plunging event in a putative cloud with mass M$_{\rm cloud}$, the interaction with the ambient medium would have significantly decreased its angular momentum. Hence, we cannot favor a formation scenario based on the observed acceleration of the centroid of the reported Br$\gamma$ line (Fig. \ref{fig:x3_spec}).
In follow-up studies, we will inspect the orbital elements of X3 and the surrounding sources in detail to search for common patterns.

\section{Conclusion} 
\label{sec:conclusion}

In this work, we used ERIS to monitor peculiar objects in the S cluster to close the gap to SINFONI IFU monitoring efforts that took place between 2005 and 2019. In particular, we have focused on G2/DSO, D9, and X7. These high-resolution adaptive optics-supported observations have a spatial pixel scale of 12.5 mas.
In addition, we included medium-resolution observations of the mini cavity and inspected the status of X3, a massive YSO located at a distance of approximately 0.1 parsec towards the supermassive black hole. In the following, we list our key findings.
\begin{enumerate}
    \item We find all S cluster sources investigated in this work at their expected positions;
    \item In comparison to the Keplerian elements that covered a data baseline until 2019, our recent measurements show accuracy of less than 2$\%$ with respect to the original orbital solutions;
    \item We successfully closed the gap between the SINFONI and ERIS observations, focusing on peculiar dusty objects;
    \item Since all the sources follow a Keplerian orbit consistent with the literature, several scenarios can be analyzed;
    \item For G2/DSO, we do not find any signs of an inspiralling trajectory leading to the destruction of the source;
    \item The D9 binary system is still intact and has not merged;
    \item The bow-shock source X7 is following its trajectory towards the north;
    \item X3, located in the mini cavity close to the evaporating cluster IRS 13, does show the same double peak Br$\gamma$ line in 2023 as observed in 2014;
    \item The centroid of the Br$\gamma$ doublet associated with X3 underwent an acceleration by about 100 km/s;
    \item Further studies will provide more details about the formation history of the massive YSO X3.
\end{enumerate}
For the future, we will strengthen our efforts to monitor peculiar sources in the inner parsec and reduce the uncertainties to less than $1\%$. This is reflected by scheduled JWST/MIRI observations or proposed observations with METIS, a first-generation instrument for the ELT operating in the MIR \citep{Brandl2021}.

\begin{acknowledgements}
We would like to thank the anonymous referee for the prompt and thorough reports that helped us improve this paper.
FP, EB, and MM gratefully acknowledge the Collaborative Research Center 1601 funded by the Deutsche Forschungsgemeinschaft (DFG, German Research Foundation) – SFB 1601 [sub-project A3] - 500700252. 
MZ acknowledges the GA\v{C}R JUNIOR STAR grant no. GM24-10599M for financial support. 
VK has been partially supported by the Collaboration Project (ref.\ GF23-04053L -- 2021/43/I/ST9/01352/OPUS 22).
VP is funded by the European Union's Horizon Europe and the Central Bohemian Region under the Marie Skłodowska-Curie Actions -- COFUND, Grant agreement \href{https://doi.org/10.3030/101081195}{ID:101081195} (``MERIT''). {VK and VP also acknowledge support from the project RVO:67985815 at the Czech Academy of Sciences.}
MM thanks the International Max Planck Research School (IMPRS) for Astronomy and Astrophysics by the Max Planck Society and the Cologne Bonn Graduate School (BCGS) for their funding.
MS is supported by the Grant Agency of Charles University under the grant number 179123.
AP, SE, JC, and GB contributed useful points to the discussion. We also would like to thank the members of the SINFONI/NACO/VISIR and ESO's Paranal/Chile team for their support and collaboration. This paper makes use of the following ALMA data: ADS/JAO.ALMA$\#$2016.1.00870.S. ALMA is a partnership of ESO (representing its member states), NSF (USA) and NINS (Japan), together with NRC (Canada), MOST and ASIAA (Taiwan), and KASI (Republic of Korea), in cooperation with the Republic of Chile. The Joint ALMA Observatory is operated by ESO, AUI/NRAO and NAOJ.
\end{acknowledgements}

\bibliography{bib}{}
\bibliographystyle{aa}

\begin{appendix}

\section{Finding chart data}

The data for the finding chart presented in Fig. \ref{fig:finding_chart} is listed in Table \ref{tab:data}.

\begin{table}[h!]
\centering
\caption{Data and bands used for the finding chart displayed in Fig. \ref{fig:finding_chart}.}

\begin{tabular}{|ccc|}
\hline
Epoch & Telescope/Instrument & Band\\
\hline
2011 & VLT/NACO    & K       \\
2008 & VLT/NACO    & L       \\
2023      & VLT/ERIS    & Br$\gamma$     \\
2016      & ALMA        & 343 GHz \\
\hline
\end{tabular}
\label{tab:data}
\end{table}

\section{Keplerian orbital solutions}

In this Appendix, we provide a complete overview of the Keplerian fit solutions of the most prominent dusty S cluster objects.

\begin{figure}[htbp!]
	\centering
	\includegraphics[width=.5\textwidth]{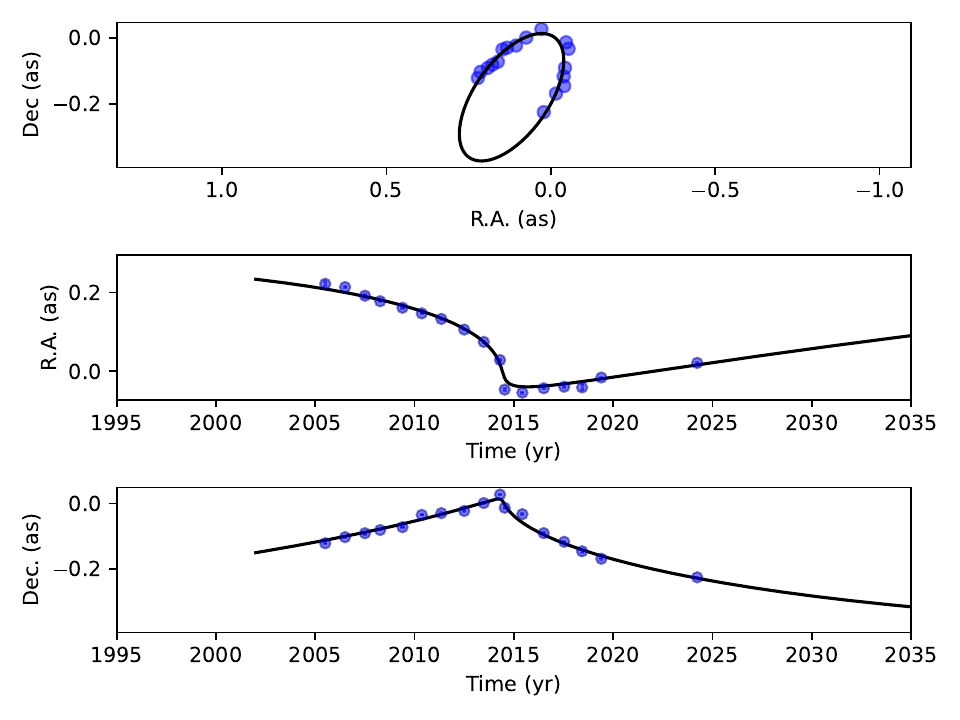}
	\caption{Keplerian fit of the astrometric positions of G2/DSO until 2024. The data includes SINFONI and ERIS observations of the Galactic center.}
\label{fig:dso_full_orbit}
\end{figure}

\begin{figure}[htbp!]
	\centering
	\includegraphics[width=.5\textwidth]{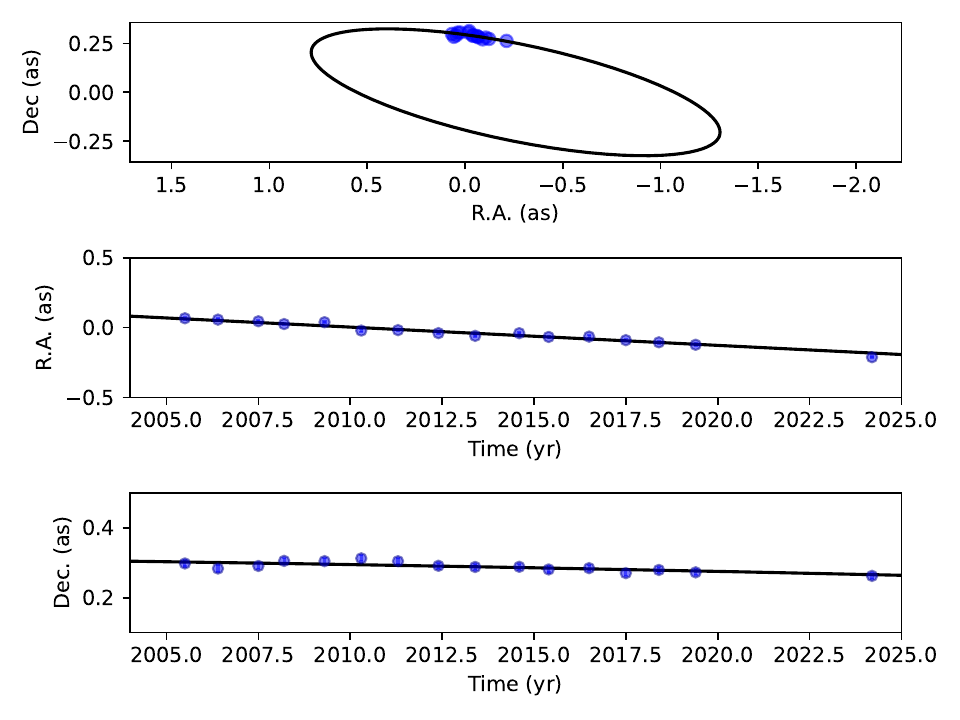}
	\caption{Orbital solution for the Keplerian approximation for D9 including observations until 2024.}
\label{fig:d9_full_orbit}
\end{figure}

\begin{figure}[htbp!]
	\centering
	\includegraphics[width=.5\textwidth]{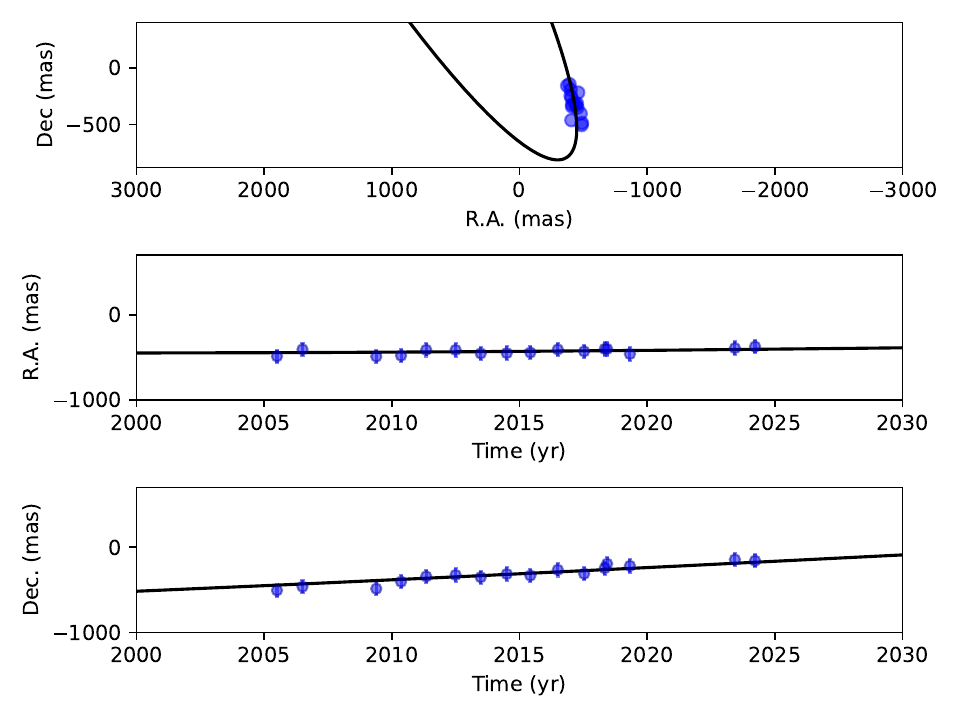}
	\caption{Evolution of the bow shock source X7. The plot shows the Keplerian fit of the astrometric positions of X7.}
\label{fig:x7_full_orbit}
\end{figure}

The orbital elements of the dusty S cluster objects are listed in Table \ref{tab:orbital_elements}. The related fit and data points are shown in Fig.~\ref{fig:dso_full_orbit}, Fig. \ref{fig:d9_full_orbit}, and Fig. \ref{fig:x7_full_orbit}. We highlight the detection of all three objects at their expected R.A. and DEC astrometric positions.

\section{Spectra of the dusty objects}

\begin{figure}[htbp!]
	\centering
	\includegraphics[width=.5\textwidth]{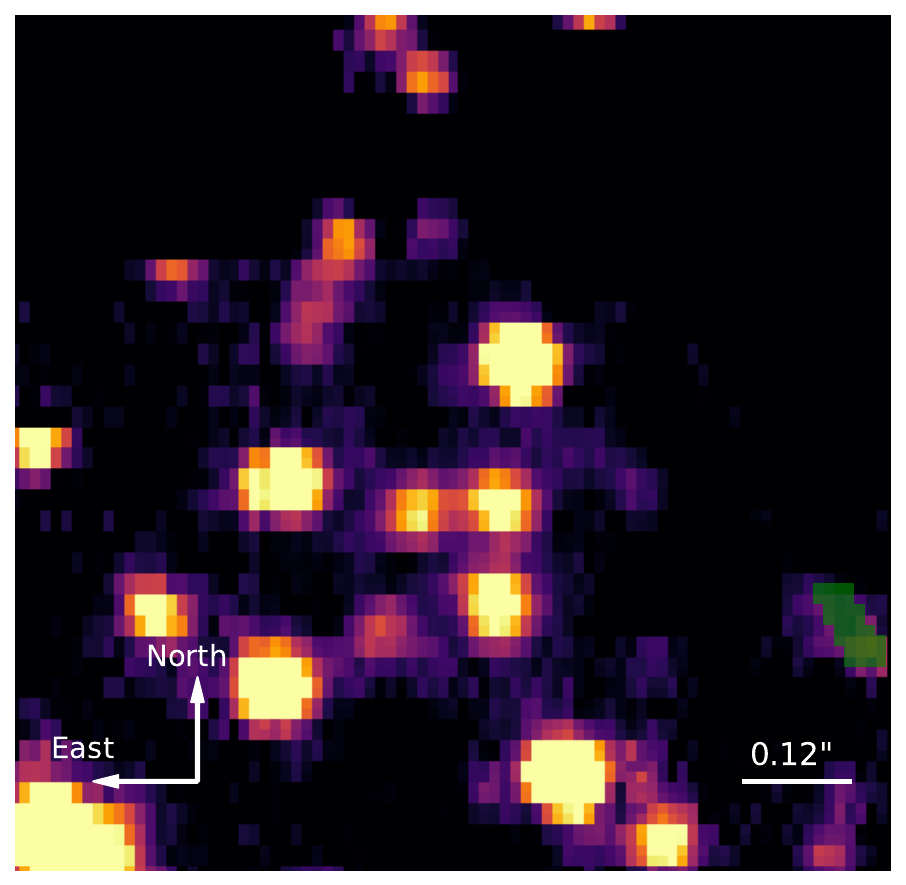}
	\caption{K-band observation of the S cluster using ERIS in 2024. On the right-hand side of the limited FOV, the bow-shock source X7 is located. The shown plot is created by selecting the IFU channels of the related Doppler-shifted Br$\gamma$ signal. Therefore, this image represents the product of the line and continuum emission of X7. From this visual inspection, we select the source and extract the spectrum that is presented in this section.}
\label{fig:extraction_example}
\end{figure}

\begin{figure*}[htbp!]
	\centering
	\includegraphics[width=1.\textwidth]{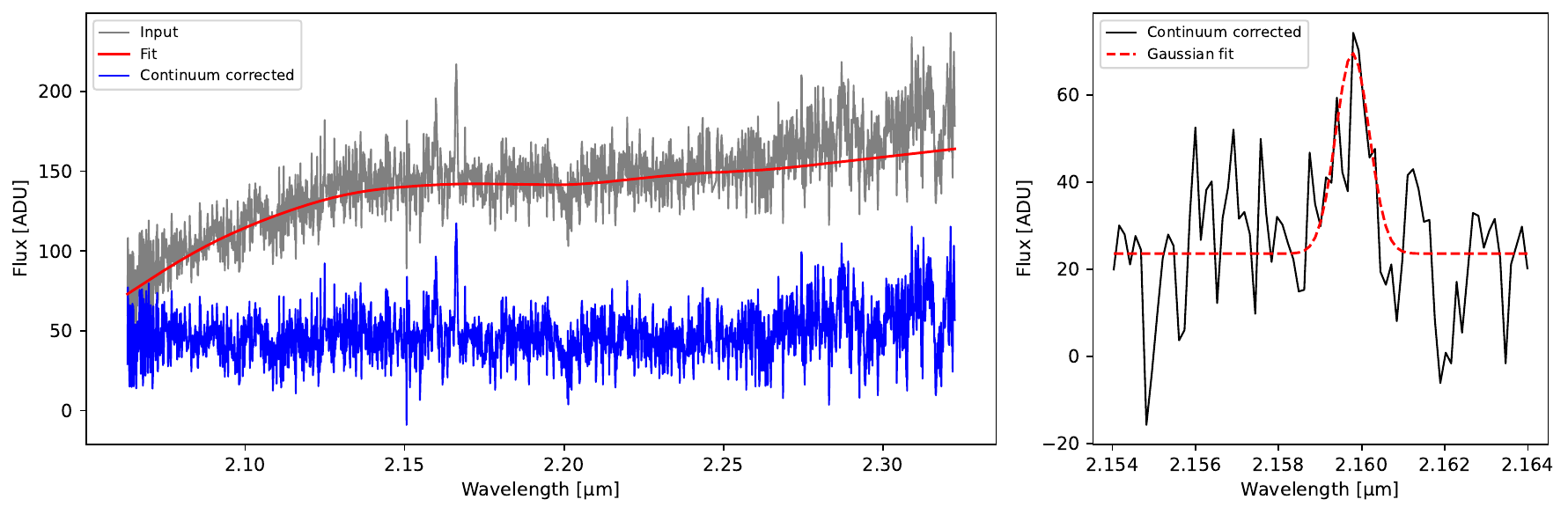}
	\caption{K-band spectrum of G2/DSO observed with ERIS in 2024. The grey spectrum represents the extracted data from the IFU cube, the blue one is corrected for the continuum slope using an Asymmetric Least Squares fit. The zoomed-in box depicts the spectral region around the Doppler-shifted peak. The Gaussian fit compared to the continuum results in a S/N ratio of 4.3.}
\label{fig:dso_spec_fit}
\end{figure*}

\begin{figure*}[htbp!]
	\centering
	\includegraphics[width=1.\textwidth]{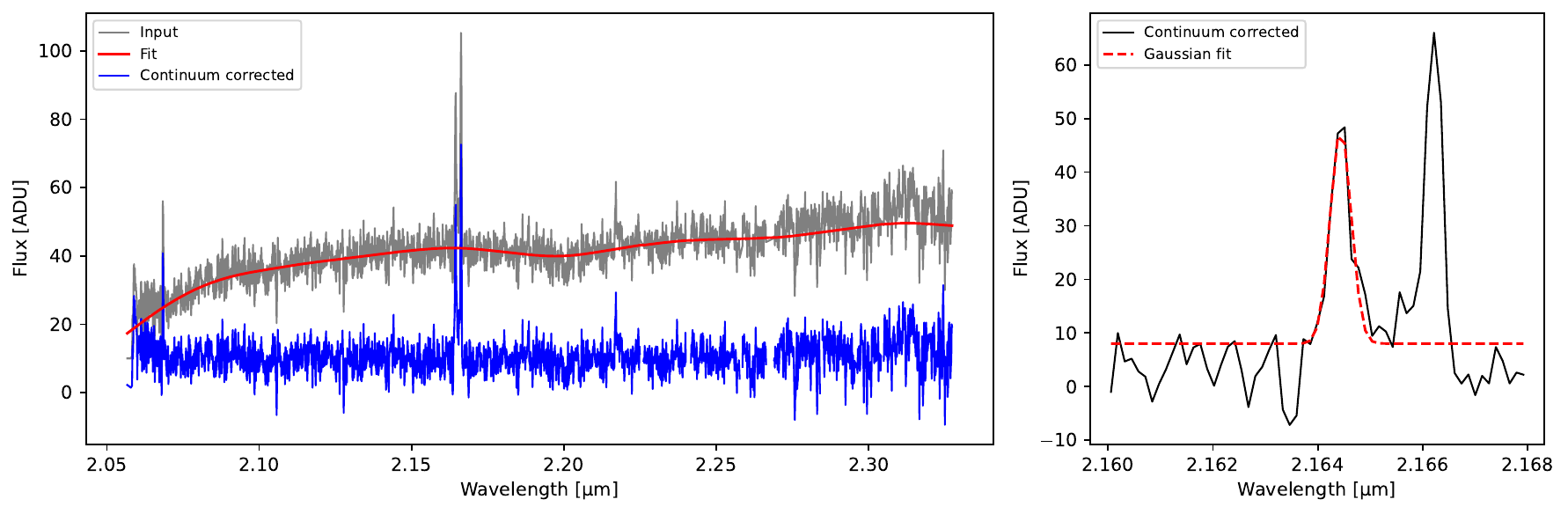}
	\caption{K-band spectrum of D9 observed with ERIS in 2024. }
\label{fig:d9_spec_fit}
\end{figure*}

\begin{figure*}[htbp!]
	\centering
	\includegraphics[width=1.\textwidth]{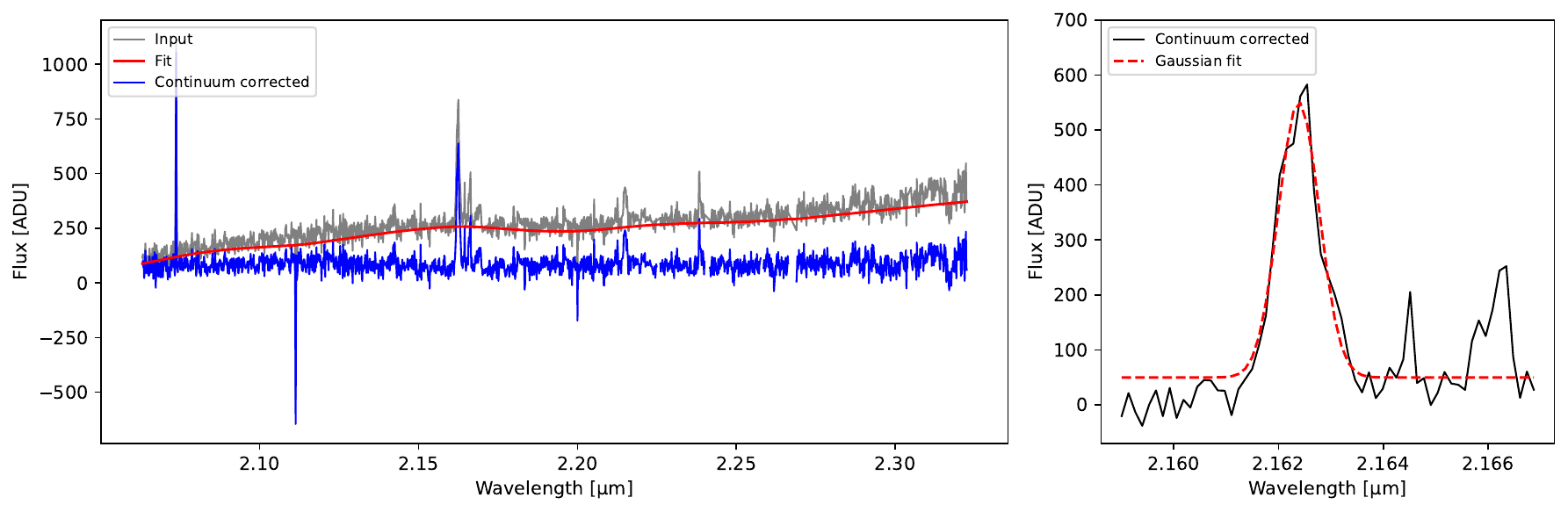}
	\caption{K-band spectrum of X7 observed with ERIS in 2024. }
\label{fig:X7_spec_fit}
\end{figure*}

In this subsection, we describe the extraction of the spectrum that results in the detection of the dusty sources analyzed in this work. The most common procedures of extracting a spectrum in the inner parsec is done using two different methods listed below.
\begin{enumerate}
    \item Direct source selection with a distant background with almost no stellar emission \citep{Eisenhauer2005}
    \item Direct source selection with using a circular-annular aperture that subtracts the closest background \citep{Valencia-S.2015} 
\end{enumerate}
Both methods listed above have their scientific justification discussed in the related publication. For the S/N ratio indicated in Table \ref{tab:data_quality}, we used neither of the methods but selected directly the dusty objects without a local or distant background subtraction. As an example, we show the process of the spectral extraction in Fig. \ref{fig:extraction_example} for X7 in the ERIS data observed in 2024.22. For this, we manually select the source of interest directly in the data and produce the spectra that are shown in Fig. \ref{fig:dso_spec_fit}-\ref{fig:X7_spec_fit}.
As mentioned in the caption of Table \ref{tab:data_quality}, we used a lower limit for the indicated S/N ratio because the measurement of this parameter is affected by the background and source selection. Another aspect, that can not be neglected, is the high stellar density in the S cluster. While the S/N ratio might be higher, Table \ref{tab:data_quality} shows conservative measurements.
In the following, we will guide reader through the process of determining the S/N ratio using the faint dusty source G2/DSO that is shown in Fig. \ref{fig:dso_spec_fit}.
As mentioned, we directly select the source using an aperture that consists of five pixels. The grey spectrum in Fig. \ref{fig:dso_spec_fit} is not background corrected and approximated with an Asymmetric Least Squares (ASL) fit shown as a red curve. The slope of this fit curve is subtracted from the grey spectrum which results in the final blue spectra (Fig. \ref{fig:dso_spec_fit}). Due to insufficient cosmic-ray correction, we manually mask all non-linear spectral features.
Finally, we use the resulting spectrum, which is corrected for the underlying continuum slope, as the basis for the S/N ratio. In Fig. \ref{fig:dso_spec_fit}, we show a zoomed-in window that represent the spectral region around the Dopplershifted Br$\gamma$ emission line of the related source.\newline
The final analyzed spectrum in shown in blue in Fig. \ref{fig:dso_spec_fit}-\ref{fig:X7_spec_fit} where we fit a Gaussian to the Doppler-shifted Br$\gamma$ emission line. {The 1$\sigma$ uncertainties of the Gaussian fit are 2.7 km/s and 3.9 km/s for X7 (Fig. \ref{fig:X7_spec_fit}) and D9 (Fig. \ref{fig:d9_spec_fit}), respectively. As discussed in Sec. \ref{sec:discuss} and mentioned in \cite{peissker2021c}, OH lines may impact the Doppler-shifted Br$\gamma$ emission measured in this work. For the spectrum of G2/DSO shown in Fig. \ref{fig:dso_spec_fit}, the Gaussian may be impacted by the OH line emission at 2.158$\mu$m \citep{Rousselot2000}, resulting in a fit uncertainty of 8.9 km/s.}
From these measurements, the S/N ratios are determined and listed in Table \ref{tab:sn_ratio_list}.
\begin{table}[htb]
\centering
\caption{Estimated S/N ratio for the dusty sources}
\begin{tabular}{cccccc}\hline \hline\\[-5pt]
             & G2/DSO & D9  & X7  \\[3pt] \hline\\[-5pt]
S/N ratio    &  4.3   & 6.2 & 13.1    \\[3pt]
\hline \hline\\[-10pt]
\end{tabular}
\tablefoot{The listed values are calculated using the Doppler-shifted amplitude of G2/DSO, D9, and X7. For the noise, we use the standard deviation of the underlying continuum in the ERIS data.}
\label{tab:sn_ratio_list}
\end{table}
We use the amplitude of the signal and divide it by the standard deviation of the continuum below and above the Dopper-shifted Br$\gamma$ peak. For the S/N listed in Table \ref{tab:data_quality}, we use conservative values of G2/DSO. Depending on the source, the S/N ratio is better as outlined in Table \ref{tab:sn_ratio_list}. We want to stress that close-by stellar sources in the crowded S cluster do have a significant impact on the quality of the extracted spectrum.

\section{Incomplete orbits}

As pointed out in \cite{ONeil2019}, the low orbital coverage of S cluster members may result in an observable-biased prior which alters the derived Keplerian approximation. In \cite{Ali2020}, we adressed this problem by dividing the stellar sample in three groups that can be distinguished by the orbital coverage of the investigated source. The first group contains sources with an orbital coverage of $40-100\%$ and for the second one, we defined a range of $20-35\%$. The dusty sources G2/DSO and D9 are in the last group with an orbital coverage of $5-15\%$, resulting in a difference of $\approx 0.6\sigma$ between the observable-biased prior and the uniform priors used in this work for the distance and mass of Sgr~A*. For X7, the situation is more complex because of the ongoing elongation of the bow shock source. While some authors prefer a core-less model that will lead to the destruction of X7 \citep{Ciurlo2023, Shaqil2025}, we have shown in this work that the bow shock source continues its path on a Keplerian orbit. While the deviation between the observed location of X7 and the inspiral orbit of Cirulo et al. increases (Fig. \ref{fig:x7_ident}), it can not be excluded that the Keplerian orbit will change dramatically due to material that is stripped away from the source. Hence, all uncertainties given in this work regarding X7 are limited to the inspected data set and may not represent the evolution of the trajectory. As we already pointed out in \cite{peissker2021}, there may have been a destructive or at least disturbing event in 2010 that may have resulted in the instability of X7. Speculatively, the presence of a binary in an optically thick envelope may explain the unusual behavior of X7 \citep{Peissker2024c}. Currently, only GRAVITY may answer the intriguing question about the possibility of a binary system inside the bright dusty L band envelope. In the future, observations with MICADO and METIS (Extremely Large Telescope) will provide a definitive answer to the question about the nature of X7. 

\section{MCMC simulations}

\begin{figure*}[htbp!]
	\centering
	\includegraphics[width=1.\textwidth]{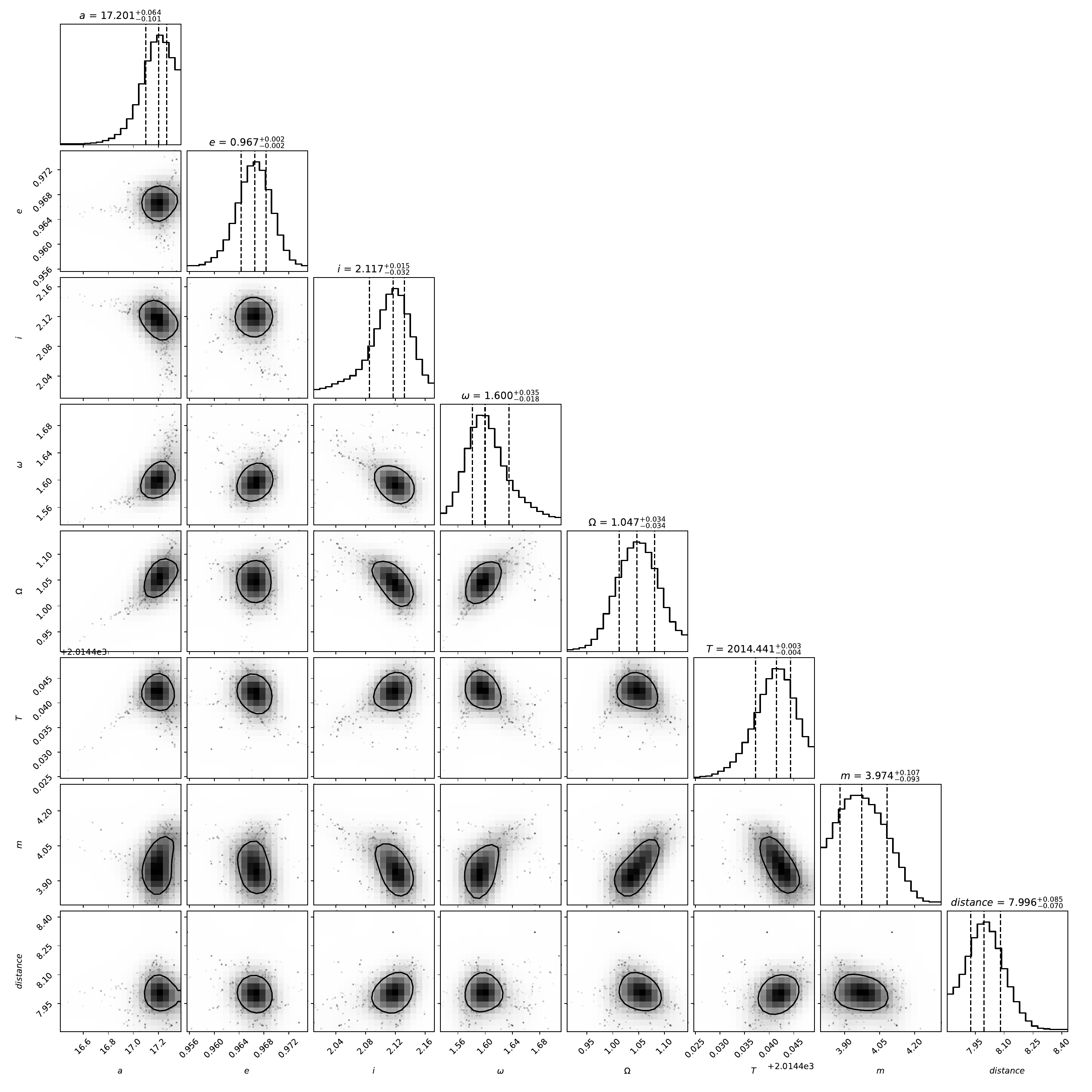}
	\caption{MCMC simulations of G2/DSO. As an input of the simulations, we use the Keplerian fit results listed in Table \ref{tab:orbital_elements}.}
\label{fig:dso_mcmc}
\end{figure*}

\begin{figure*}[htbp!]
	\centering
	\includegraphics[width=1.\textwidth]{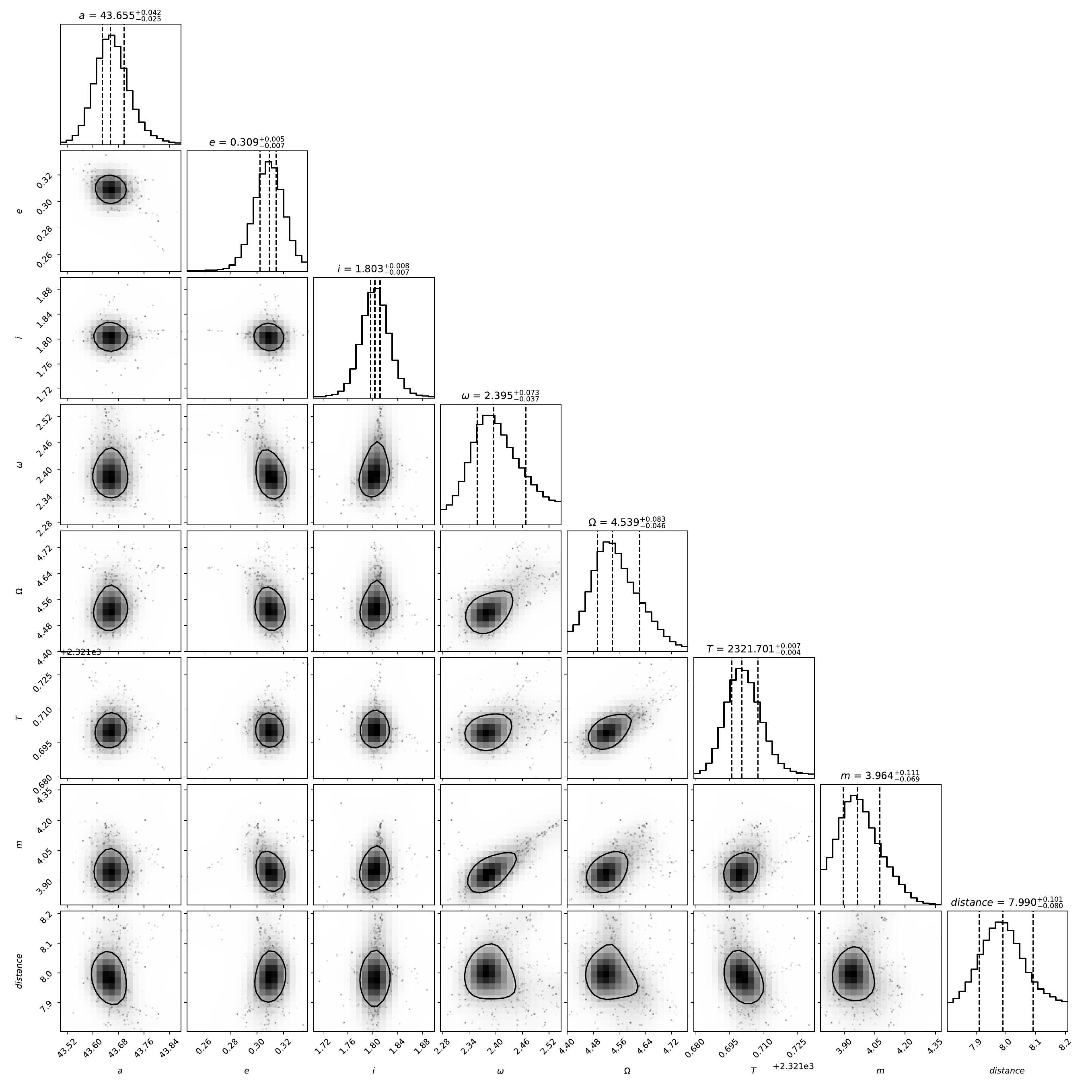}
	\caption{MCMC simulations of D9.}
\label{fig:d9_mcmc}
\end{figure*}

\begin{figure*}[htbp!]
	\centering
	\includegraphics[width=1.\textwidth]{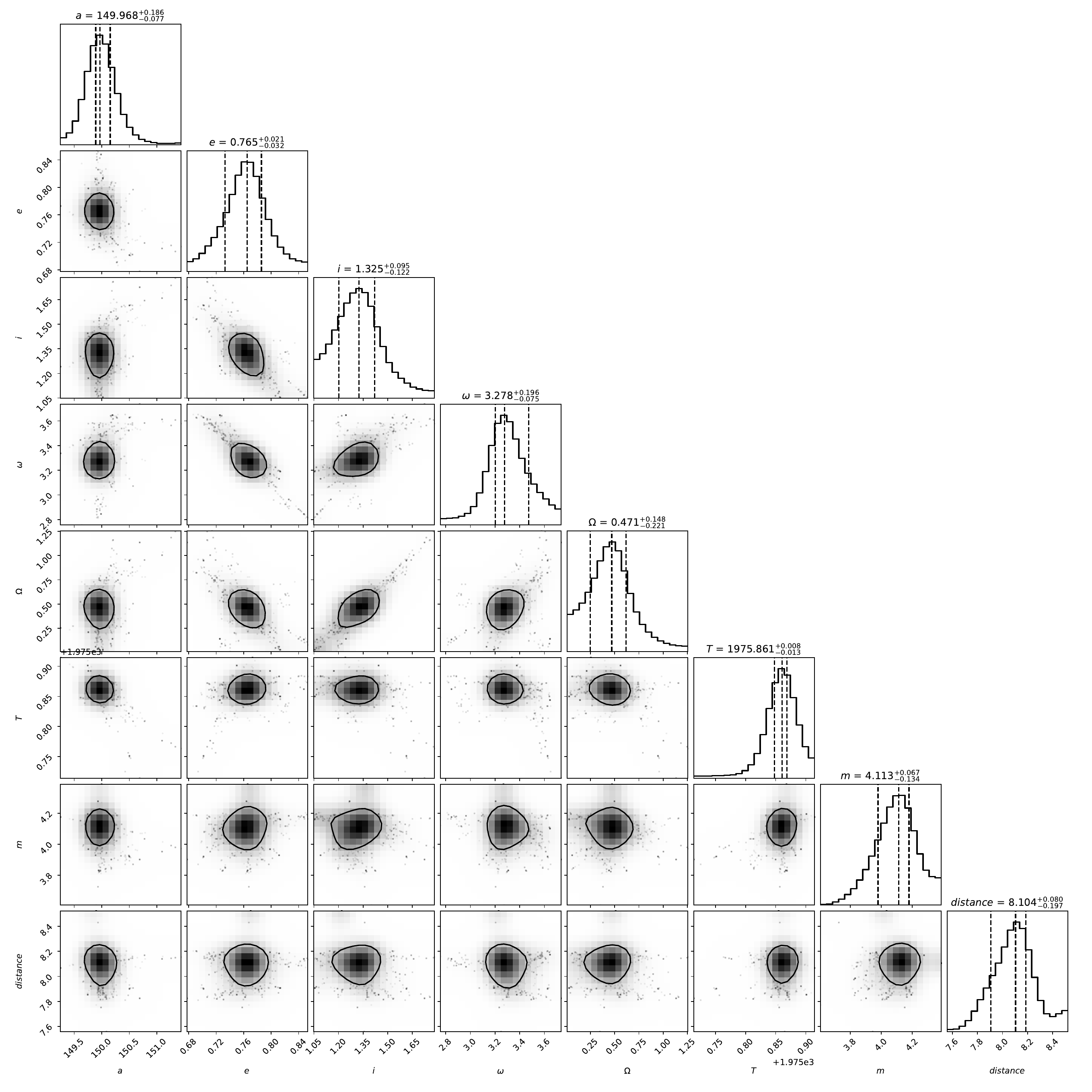}
	\caption{MCMC simulations of X7.}
\label{fig:x7_mcmc}
\end{figure*}

The MCMC simulations were performed in a similar way as it was done in \cite{Peissker2020d, peissker2021c, Peissker2022, peissker2023c, Peissker2024a}. We use the results of the Keplerian fit as the apriori distribution. From the fits presented in Fig. \ref{fig:dso_mcmc}-\ref{fig:x7_mcmc} we extract the 1$\sigma$ posterior uncertainty distribution. Since the uncertainty range seems slightly asymmetric with a marginal lower limit, we use the upper limit to cover aspects of the data that may be underrepresented in the fit. As pointed out by \cite{ONeil2019}, the orbital coverage of some S cluster members is low compared to their complete trajectory. Due to different background scenarios and nearby stars, any object in the S cluster suffer from multiple sources of interference that are not reflected in the here presented uncertainty range. This problem becomes evident by looking at the simulated expected observations with MICADO, a future instrument at the ELT \cite{Trippe2010}. The background in the S cluster may be dominant for fainter sources which potentially lead to astrometric deviations outside of the uncertainty range listed in \ref{tab:orbital_elements} or discussed in \cite{ONeil2019}. We refer the interested reader to \cite{Peissker2022} for a discussion about the chances of observing stars close to the detection limit. Furthermore, we want to stress that the impact of all sources of influence decreases with an increasing data baseline.  

\end{appendix}

\end{document}